\documentclass[reprint,nofootinbib,superscriptaddress,amsmath,amssymb,aps,prd]{revtex4-2}
\usepackage{amsmath,mathrsfs,amsbsy,color,graphicx,bm,amsthm,amsfonts}
\usepackage{bbm}
\usepackage{times}
\usepackage{units}
\usepackage{dcolumn}
\usepackage{graphicx}
\usepackage{epsfig}
\usepackage{epstopdf}
\usepackage[colorlinks,linkcolor=red,anchorcolor=green,citecolor=blue,CJKbookmarks=True]{hyperref}
\DeclareMathSymbol{\shortminus}{\mathbin}{AMSa}{"39}
\usepackage{mathrsfs}
\usepackage{braket}
\usepackage{amssymb}
\usepackage{txfonts}
\usepackage{float}
\usepackage{enumitem}
\usepackage{multirow}

\newcommand{\meq}[1]{(\ref{#1})}

\newcommand{\pp}{\partial}

\begin{document}
\title{Lorentz violation signatures in the low-energy sector of Ho\v{r}ava gravity from black hole shadow observations}

\author{Wentao Liu}
\thanks{These authors contributed equally to this work.}
\affiliation{Department of Physics, Key Laboratory of Low Dimensional Quantum Structures and Quantum Control of Ministry of Education, and Synergetic Innovation Center for Quantum Effects and Applications, Hunan Normal
University, Changsha, Hunan 410081, P. R. China}

\author{Hongxia Huang}
\thanks{These authors contributed equally to this work.}
\affiliation{Department of Physics, Key Laboratory of Low Dimensional Quantum Structures and Quantum Control of Ministry of Education, and Synergetic Innovation Center for Quantum Effects and Applications, Hunan Normal
University, Changsha, Hunan 410081, P. R. China}

\author{Di Wu}
\email[]{wdcwnu@163.com} \affiliation{School of Physics and Astronomy, China West Normal University, Nanchong, Sichuan 637002, P. R. China}

\author{Jieci Wang}
\email[]{jcwang@hunnu.edu.cn} \affiliation{Department of Physics, Key Laboratory of Low Dimensional Quantum Structures and Quantum Control of Ministry of Education, and Synergetic Innovation Center for Quantum Effects and Applications, Hunan Normal
University, Changsha, Hunan 410081, P. R. China}

\begin{abstract}

In this paper, we use the Ho\v{r}ava gravity model and EHT observations of supermassive black holes (BHs) to investigate signatures of Lorentz violation in real astrophysical environments.
The Lorentz violation in the rotating Ho\v{r}ava BH spacetime are confined to the strong gravitational field region, being induced by the BH's rotation. 
Due to the non-separability of the photon motion equations in this spacetime, we employed a numerical backward ray-tracing method to generate shadow images for various BH parameters. 
Subsequently, we extracted coordinate positions characterizing the shadow shape from high-pixel images to evaluate the parameter space of the BH.
When evaluating M87*, Lorentz violation can occur with arbitrary strength. However, for Sgr A*, we can impose certain parameter constraints on Lorentz violation.  These constraints depend on the BH's spin.  If future observations confirm Sgr A*'s spin parameter less than 0.81 at maximum inclination, current EHT results would challenge general relativity and support Lorentz violation in low-energy regimes. 
\end{abstract}

\maketitle

\section{Introduction}\label{Sec.1}
Over the past two decades, extensive theoretical research has explored the violation of Lorentz symmetry through various models inducing anisotropic spacetime \cite{Kostelecky1998,Horava:2009uw,Aharony:1999ti,Kalb:1974yc}, as part of the broader pursuit of a quantum gravity theory. 
These models disrupt Lorentz invariance at the Planck scale, yet the resulting spacetime effects remain challenging to detect at significantly lower energy levels.
Models such as the Einstein-Bumblebee \cite{Kostelecky2001,Casana2018,Maluf2021,Filho:2022yrk,Liu:2022dcn,AraujoFilho:2024ykw,Liu:2024wpa,Liu:2025bpp,Chen:2025ypx} and Kalb-Ramond \cite{Yang:2023wtu,Duan:2023gng,Filho:2023ycx,Liu:2024oas,Liu:2024lve} gravity theories, which have garnered significant recent attention, feature black hole (BH) spacetimes that deviate from Minkowskian behavior at infinity. 
Consequently, solar system observations effectively limit the extra degrees of freedom tied to Lorentz violation in these models to minimal values \cite{Yang:2023wtu} and render their astrophysical signatures, such as those in gravitational waves \cite{LIGOScientific:2016aoc} or BH images \cite{EventHorizonTelescope:2019dse}, so delicate that current detection techniques can hardly verify them.
In a recent and intriguing study, Devecio\v{g}lu and Park employed a straightforward two-step method to derive exact Kerr-type solutions within the low-energy regime of Ho\v{r}ava gravity \cite{Devecioglu:2024uyi}, which is a four-dimensional theory that exhibits spontaneous Lorentz symmetry breaking \cite{Park:2023byp}.
In this BH spacetime, Lorentz violation is caused by the combined effects of a strong gravitational field and frame-dragging. 
The behavior at infinity remains asymptotically flat, thus indicating that Lorentz violation is primarily evident near the BH horizon. 
It is well-known that detecting Lorentz violation in an asymptotically flat region at large distances from the BH is challenging. 
However, in this spacetime, detecting Lorentz violation becomes more feasible in the strong gravitational field around a BH. 
Observation of BH shadows, as the most direct method to understand strong gravitational fields, is also the most likely macroscopic field in which Lorentz violation can be detected.

Inspired by the pioneering observations from the EHT \cite{EventHorizonTelescope:2019uob,EventHorizonTelescope:2019jan,EventHorizonTelescope:2022wkp,EventHorizonTelescope:2022apq} and combining the low-energy regime of Ho\v{r}ava gravity models, we can search for signatures of Lorentz violation in the observational data.
Picture a coin placed beneath a sheet of paper. 
Even though the coin itself is invisible, its shape and features remain unchanged, and each time you rub the paper, the impression consistently reveals the same traits.
In a similar way, the spacetime structure of a BH is fixed. 
Regardless of the shape, color, or behavior of external electromagnetic radiation, the BH's influence stays constant provided a few basic rules are upheld. 
The presence of the BH can be revealed through illumination from external electromagnetic sources.
Under this ``light", the BH shadow, a unique characteristic defined by its shape and size, will thus be captured, potentially revealing distinct astronomical signals and information shaped by strong gravitational effects, including possible evidence of Lorentz violation in astronomical environments \cite{Song:2025myx,Finke:2024ada,Wei:2016exb}.
By comparing the differences between theoretical and observational results, we can search for possible signatures of Lorentz violation in the observational data.

\section{Models}\label{Sec.2}
The action describing the low-energy sector of (non-projectable) Ho\v{r}ava gravity, up to boundary terms, is given by \cite{Horava:2009uw}
\begin{equation}
\mathcal{S}_g=\int_{\mathbf{R}\times\Sigma_t}dtd^3x\sqrt{g}N\left[\frac{1}{\kappa}\left(K_{ij}K^{ij}-\lambda K^2\right)+\xi R \right],
\end{equation}
where $ K_{ij}=(2N)^{-1}(\dot{g}_{ij}-\nabla_iN_j-\nabla_jN_i) $ is the extrinsic curvature and $ R $ is the three-curvature in the ADM metric.
Last year, Park et al. found exact solutions as follows \cite{Devecioglu:2024uyi}
\begin{equation}\label{ds1}
ds^2_1=-N^2dt^2 +\frac{\rho^2}{\Delta_r}dr^2+\rho^2d\theta^2 +\frac{\Sigma^2\sin^2\theta}{\rho^2}\left(d\varphi+N^\varphi dt\right)^2.
\end{equation}
Here, the following definitions are introduced
\begin{eqnarray*}
&\Sigma^2=\left(r^2+a^2\right)\rho^2+f(r)a^2\sin^2\theta,\quad N^\varphi=-\frac{g(r)}{\Sigma^2},\\
&\rho^2=r^2+a^2\cos^2\theta,\qquad\qquad\qquad N^2=\frac{\rho^2\Delta_r(r)}{\Sigma^2},
\end{eqnarray*}
with three functions that depend on $ r $
\begin{equation*}
f(r)=2Mr,~~~ g(r)=2aMr\sqrt{\kappa\xi},~~~ \Delta_r(r)=r^2+a^2-2Mr,
\end{equation*}
which reduce to the known Kerr solution in the GR case of $ \xi=1/\kappa $ or to the Schwarzschild solution when $ a=0 $.
The $ M $ is an integration constant and can be regarded as the mass parameter of the BH.
It is straightforward to observe the non-trivial Lorentz-violating effect in the $g_{tt}$ and $g_{t\varphi}$ components when $ \xi \neq 1/\kappa $.
An additional noteworthy property is that the solution \meq{ds1} remains valid for any arbitrary $ \lambda $ because $ K=0 $, indicating 'maximal' slicing.

To better understand the differences between this solution and the Kerr solution, we can set $ \ell\equiv\sqrt{\kappa\xi}-1 $ and rescale the line element $\meq{ds1}$, as follows
\begin{equation}\label{ds2}
ds^2_2=ds^2_\mathrm{Kerr}+\frac{4\ell aMr\sin^2\theta}{\rho^2}\left(\frac{\ell+2}{\Sigma^2}aMrdt^2-dtd\varphi\right),
\end{equation}
where $ ds^2_\mathrm{Kerr} $ is represents the line element of the Kerr BHs.
We can refer to $ \ell $ as the Lorentz-violating parameter. 
When it is zero, which corresponds to $ \xi=1/\kappa $, the correction terms due to Lorentz violation disappear, and spacetime reverts to the Kerr spacetime.
Moreover, the correction terms vanish in the presence of a non-rotating BH and therefore, no corresponding static Lorentz-violating BH exists.
In Boyer-Lindquist coordinates, there are two horizons at $ r_{\pm} $, where $ g^{rr}=\Delta_r/\rho^2=0 $, consistent with the Kerr spacetime. 
This condition is satisfied when $ r^2+a^2-2Mr=0 $, thus, the horizons are $ r_\pm=M\pm\sqrt{M^2-a^2} $.

When a BH is positioned between an observer and an extended background light source, only some of the photons emitted by the source and deflected by the BH's gravitational field reach the observer. 
The apparent shape of the BH, represented by its shadow contour, is closely related to the photon geodesics in the BH's spacetime.
The Hamiltonian of photon motion along a null geodesic in viable Lorentz-violating spacetime can be expressed as 
\begin{equation}\label{EqHJ}
\mathcal{H}=\frac{1}{2}g^{\mu\nu}p_{\mu}p_{\nu}.
\end{equation}
Here, $ g^{\mu\nu} $ is the metric associated with the line element \eqref{ds2}, and $ p_\mu $, the dual of the photon's 4-momentum vector, is given by $ p_\mu \equiv g_{\mu\nu} p^\nu $. 
The 4-momentum vector of a photon is written as
\begin{equation}
p^\nu=\left(\dot{t},\dot{r},\dot{\theta},\dot{\varphi}\right),
\end{equation}
where the dot denotes the derivative with respect to the affine parameter $ \tau $.
By setting the Killing fields $ \xi^\mu=(\partial/\partial t)^\mu $ and $ \psi^\mu=(\partial/\partial \varphi)^\mu $, the two conserved quantities in the spacetime, the energy $ \mathcal{E} $ and the angular momentum along the axis of symmetry $ L_z $, can be defined as \cite{Waldbook,Zeng:2021mok,Liu2023}:
\begin{equation}
\mathcal{E}=-g_{\mu\nu}\xi^\mu\dot{x}^\nu,\quad\quad
L_z=g_{\mu\nu}\psi^\mu\dot{x}^\nu,
\end{equation}
respectively.
With these two conserved quantities, we can obtain all the null geodesic equations governing photon trajectories, the specific form of which is given in Appendix \ref{AppendixA}.
These equations determine the behavior of light propagation within the spacetime influenced by rotating BHs.
Obviously, hamiltonian constraint condition \meq{HHHHH} shows that the Lorentz-violating parameter $ \ell $ prevents the null geodesic equations from being variable-separable. 
This occurs due to the absence of a Carter-like constant \cite{Carter:1968rr} in the rotating spacetime \meq{ds2}, leaving only two integrals of motion, $ \mathcal{E} $ and $ L_z $, in this case.
For non-separable systems, we cannot obtain the radial equation for photons and its effective potential, which prevents us from deriving the parametric equations for the shadow profile by defining the impact parameter \cite{Perlick:2021aok}.
Therefore, we employ the numerical backward ray-tracing technique \cite{Cunha:2015yba,Hu:2020usx,Zhong:2021mty} to trace the paths of null geodesics in reverse time from the observer.
Specifically, the data transmitted by each null geodesic is projected onto a corresponding pixel in the final image on the observer's image plane.
Then, for each pixel, we solve the null geodesic equations numerically, incorporating specific initial conditions.
Thus, the BH's shadow image is formed by pixels linked to null geodesics that cross the event horizon.
The detailed procedure is provided in Appendix \ref{AppendixB}.

\section{Results}
We now analyze the impact of the Lorentz-violating parameter $\ell$ on the BH shadows.
\begin{figure}[h]
\centering
\includegraphics[width=0.9\linewidth]{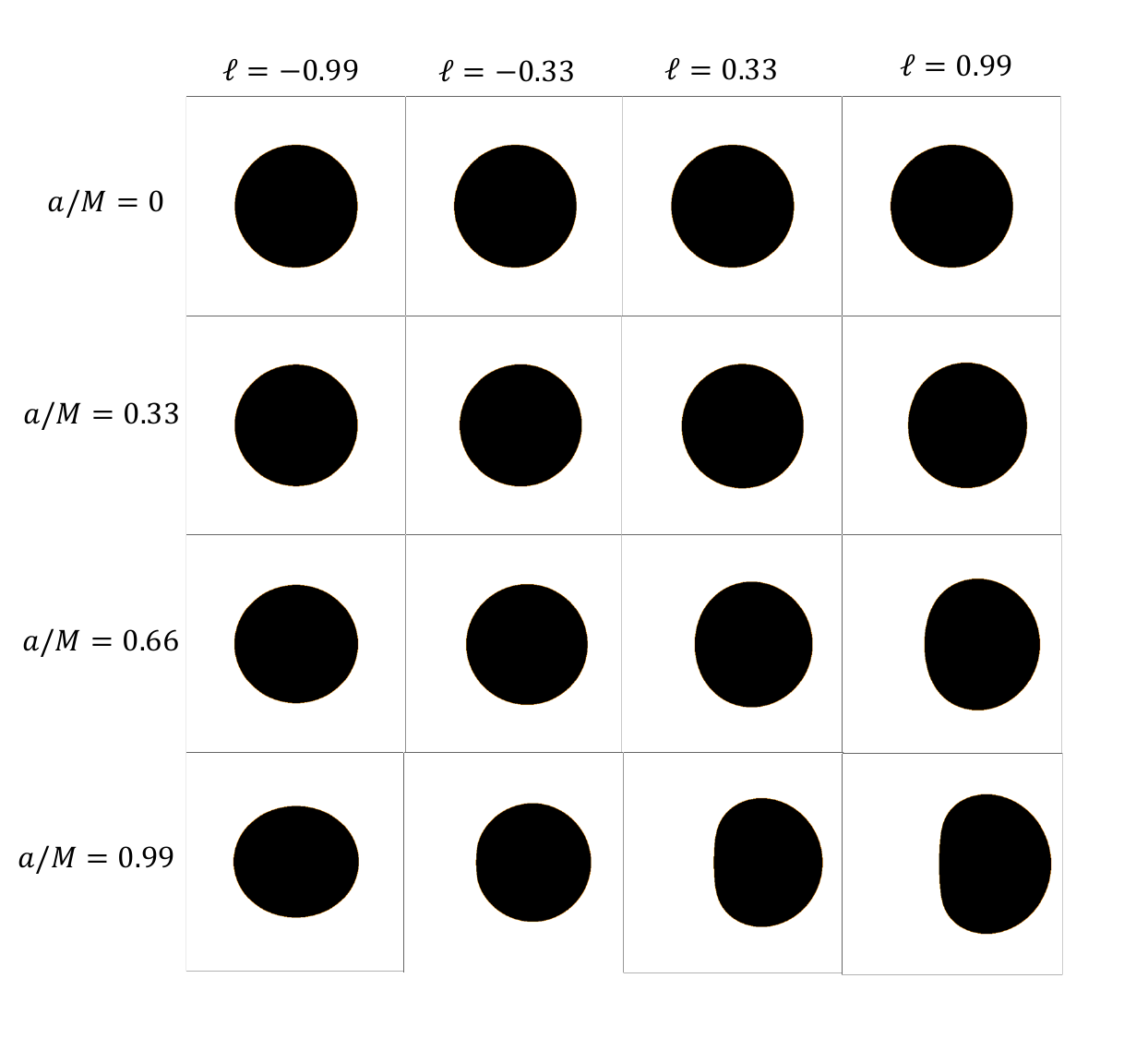}
\caption{Shadows cast by rotating Ho\v{r}ava BHs, as seen by an observer at $ \theta_0=\pi/2 $.}
\label{fig1n}
\end{figure}
Fig. \ref{fig1n} illustrates the shadows cast by various parameters at an observational inclination of $\pi/2$.
Here, we consider the parameter range $\ell \in \left(-1, 1\right)$, corresponding to $ \sqrt{\kappa \xi} $ ranging from 0 to 2, which is physically reasonable.
For a non-rotating BH, all images appear identical, as directly inferred from the spacetime metric. 
For a rotating BH, we need to consider separately the cases of $\ell>0$ and $\ell<0$, because they exhibit completely different trends.
With $ \ell>0 $, the parameter does not directly change the size of the BH shadow; instead, as the parameter increases, the contour becomes increasingly flattened. 
However, this does not imply a degeneracy between the rotational parameter and the Lorentz-violating parameter. 
In fact, the Lorentz-violating parameter causes the shadow contour to transition from a circular shape to an O-like shape, while the rotational parameter causes the shadow contour to transition from a circular shape to a D-like shape.
Although strong Lorentz violation causes the silhouette of a rotating BH to resemble that of an extremal rotating BH in GR, the BH parameters can still be inferred from its characteristic quasi-linear boundary.
With $ \ell<0 $, some very interesting phenomena emerge. 
At extreme spin parameters, as negative Lorentz violation increases, the contour first becomes nearly a perfect circle before flattening along the y-axis and eventually forming a horizontally oriented O-shape.
Interestingly, under reasonable parameter configurations, a BH can exhibit a shadow similar to the circular shape of a static BH with frame-dragging inside the lensing ring.
In Appendix \ref{AppendixC}, we present the complete information on light rays around the BH, including more parameters and the lensing ring.
These unique combinations provide potential evidence for exploring Lorentz violation in strong gravitational fields through observations with the EHT.

To effectively characterize the shadow, it is crucial to use measurable and reliable properties.
We utilize two such observable measurements: the radius $ R_s $ and the distortion parameter $ \delta_s $, as outlined in Ref. \cite{Hioki:2009na}.
\begin{figure}[h]
\centering
\includegraphics[width=0.75\linewidth]{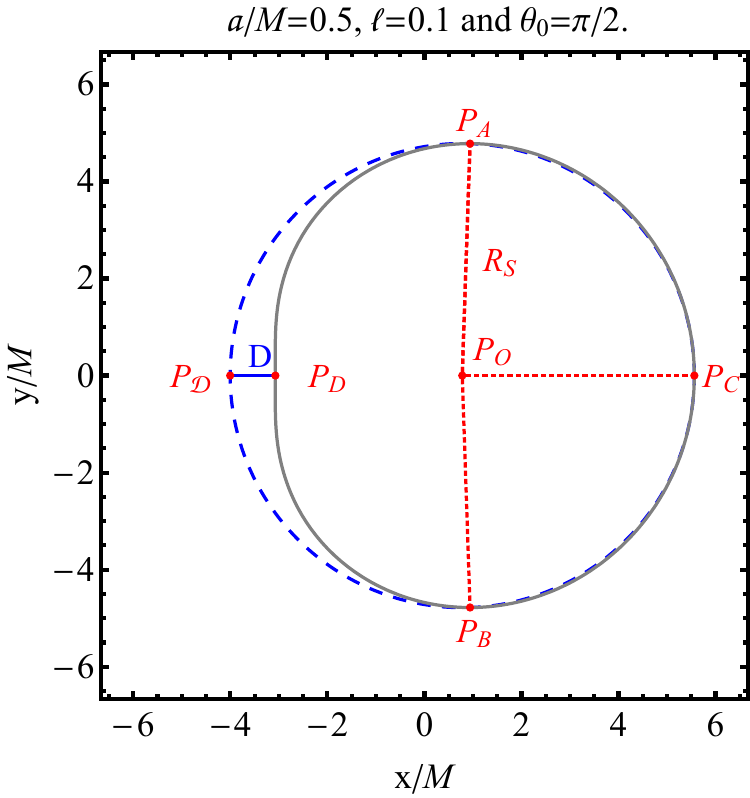}
\caption{The observable measurements for the apparent shape of the BH are the radius of the reference circle $ R_s $ and the distortion parameter $ \delta_s=\text{D}/R_s $, where $ \text{D} $ is the difference between the left endpoints of the reference circle and of the shadow.}
\label{fig7}
\end{figure}
The size of the shadow is characterized by the parameter $ R_s $, which corresponds to the radius of the reference circle shown as a blue dashed line in Fig. \ref{fig7}.
Passing through three points, the reference circle intersects the top position $P_A=(x_t,y_t)$, the bottom position $P_B=(x_b,y_b)$, and the point $P_C=(x_r, 0)$.
The point $P_C$ corresponds to the unstable retrograde circular orbit as seen from the equatorial plane by an observer.
Additionally, the point $ P_O=(x_o,0) $ represents the center of the reference circle, and $ x_o $ can be derived from the coordinates of $ P_A, P_B $, and $ P_C $:
\begin{equation}
x_o=\frac{x^2_r-x^2_t-y^2_t}{2(x_r-x_t)}.
\end{equation}
Next, the difference between the shaded left-hand point $ P_D=(x_d,0) $ and the reference circle's left-hand point $ P_\mathcal{D}=(\tilde{x}_r,0) $ needs to be considered.
The size of this difference is evaluated by $ \mathrm{D}=|x_d-\tilde{x}_r|$.
Further, the two observable measurements are defined as follows \cite{Tang:2022uwi}:
\begin{equation}
R_s=\frac{(x_t-x_r)^2+y_t^2}{2|x_r-x_t|}, \quad\quad \delta_s=\frac{\mathrm{D}}{R_s}.
\end{equation}
\begin{figure}[h]
\centering
\includegraphics[width=0.45\linewidth]{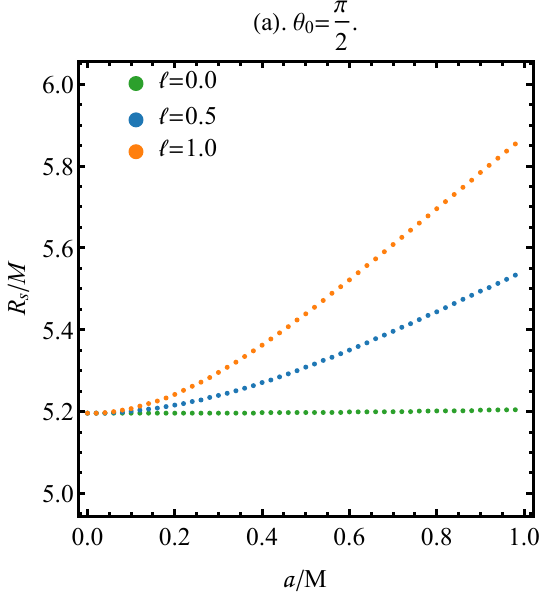}
\includegraphics[width=0.45\linewidth]{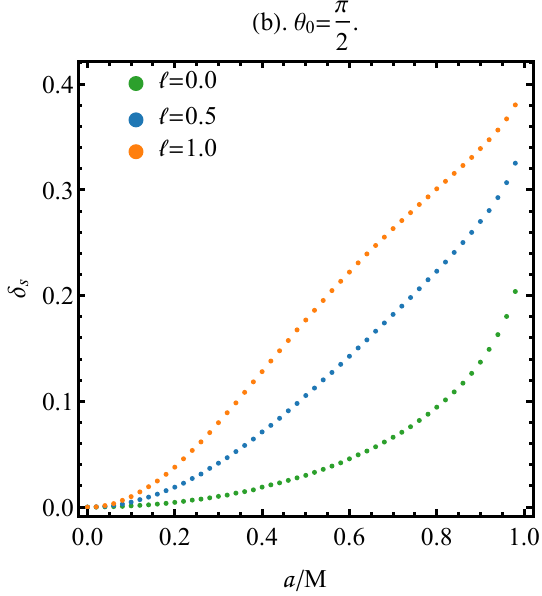}
\includegraphics[width=0.45\linewidth]{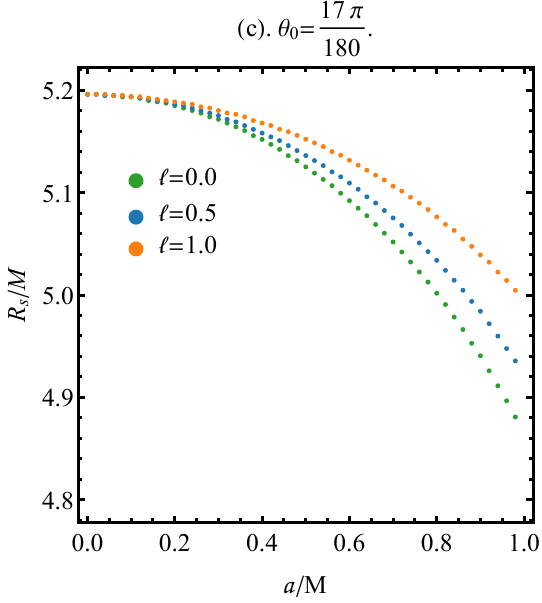}
\includegraphics[width=0.455\linewidth]{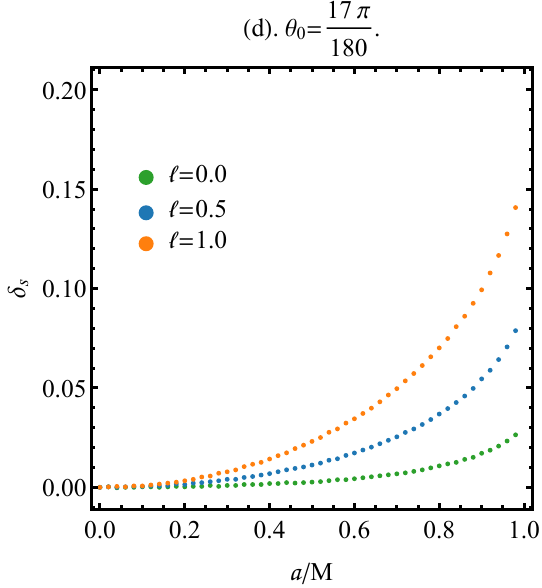}
\caption{Observable measurements $ R_s $ (left) and $ \delta_s $ (right) as functions of $ a/M $ are presented, respectively.}
\label{fig8}
\end{figure}
We numerically calculated these two observable measurements with the observer inclination angle fixed at $ \theta_0=\pi/2 $ or $ \theta_0=17\pi/180 $, where the latter corresponds to the actual angle of observation of M87* from Earth.
The results are presented in Fig. \ref{fig8}.
Given that a non-integrable system does not have an analytical shadow contour edge, we extract shadow edge pixels from the BH image with the corresponding parameters. 
This allows us to determine the positions corresponding to $ P_{A,B,C,D} $ in Fig. \ref{fig7}. 
Specifically, each point in Fig. \ref{fig8} corresponds to a 16K-pixel shadow image to achieve sufficient precision. Moreover, the interval between each pair of points is 0.02, and the spin parameter ranges from 0 to 0.98. 
It should be noted that when the BH does not rotate, the effects due to Lorentz violation and the BH's gravitational effects are decoupled, such that under all Lorentz-violating parameters $ \ell $, the starting points of the visualization curves as the spin parameter $ a/M $ increases correspond to the Schwarzschild BH.
Fig. \ref{fig8}(a) and Fig. \ref{fig8}(c) show the effect of the spin parameter on $ R_s $. 
Within the range of the spin parameter $ a/M<0.1 $, the effects of the Lorentz-violating parameter on $ R_s $ are nearly indistinguishable in the images. 
Interestingly, for observer inclination angle $ \theta_0=\pi/2 $ , when the spin parameter exceeds 0.3, the relationship between the spin parameter and $ R_s $ under different Lorentz violations exhibits a quasi-linear behavior.
Note that although the line corresponding to $ \ell=0 $ is not parallel to the coordinate axis, the effect of the BH spin on $ R_s $ under this parameter is very small.
In Fig. \ref{fig8}(c), it can be observed that, at smaller inclination angles, Lorentz violation can lead to a larger angular radius of the shadow in rotating BHs.
This implies that, in the context of Lorentz violation, the observational data of M87* can actually be compatible with larger BH spin parameters.
Figs. \ref{fig8}(b) and \ref{fig8}(d) clearly show the distortion parameter $ \delta_s $ of the shadow. 
Regardless of the inclination angle, Lorentz violation consistently increases the degree of shadow distortion. 
For larger inclination angles, for example, $ \theta_0=\pi/2 $, this effect can also alter the trend of shadow contour distortion caused by BH spin, shifting it from an accelerating increase with spin to a decelerating one.


The pioneering observation of a BH shadow, achieved by the Event Horizon Telescope collaboration, has greatly advanced our knowledge of these remarkable entities. 
First, we will explore the potential properties of this astronomical entity by utilizing the near-circular shape and dimensions of the M87* shadow, and examine the hypothesis that M87* could be a Ho\v{r}ava BH, ensuring adherence to the Kerr bound $ a/M<1 $. 
For an approximate estimation, we employ the metric \meq{ds2} in Ho\v{r}ava theory to determine the angular radius of a BH shadow, expressed as $ \theta_{\text{BH}} = R_s \frac{\mathcal{M}}{D_O} $, where $D_O$ denotes the observer-BH distance. 
In particular, considering a BH with a mass $ \mathcal{M} $ and located $ D_O $ from the observer, we quantify the angular radius $ \theta_\text{BH} $ as $ \theta_\text{BH} = 9.87098 \times 10^{-6} R_s \left(\frac{\mathcal{M}}{M_\odot}\right) \left(\frac{1 \text{kpc}}{D_O}\right) \mu\text{as} $ \cite{Amarilla:2011fx}. 
In the case of the M87* BH, its mass amounts to $\mathcal{M} = 6.5 \times 10^9 M_\odot$, and the distance to the observer is $D_O = 16.8 Mpc$ \cite{EventHorizonTelescope:2019ggy}. 
The observational data from the EHT show that the angular diameter of the M87* BH should be constrained within the range of $42 \pm 3.0 \mu\text{as}$ \cite{EventHorizonTelescope:2019dse,Long:2019nox,Pal:2023wqg}.
We searched the parameter space $ (\ell,a/M) $ of a Ho\v{r}ava BH under the observational inclination angle $ \theta_0=17\pi/180 $. 
We numerically calculated the parameter grid for $ \ell\in[0,1) $ and $ a/M\in[0,1) $ with a parameter interval of $ 0.01 $, and plotted the parameter space map that satisfies the EHT observational requirements.
\begin{figure}[h]
\centering
\includegraphics[width=0.8\linewidth]{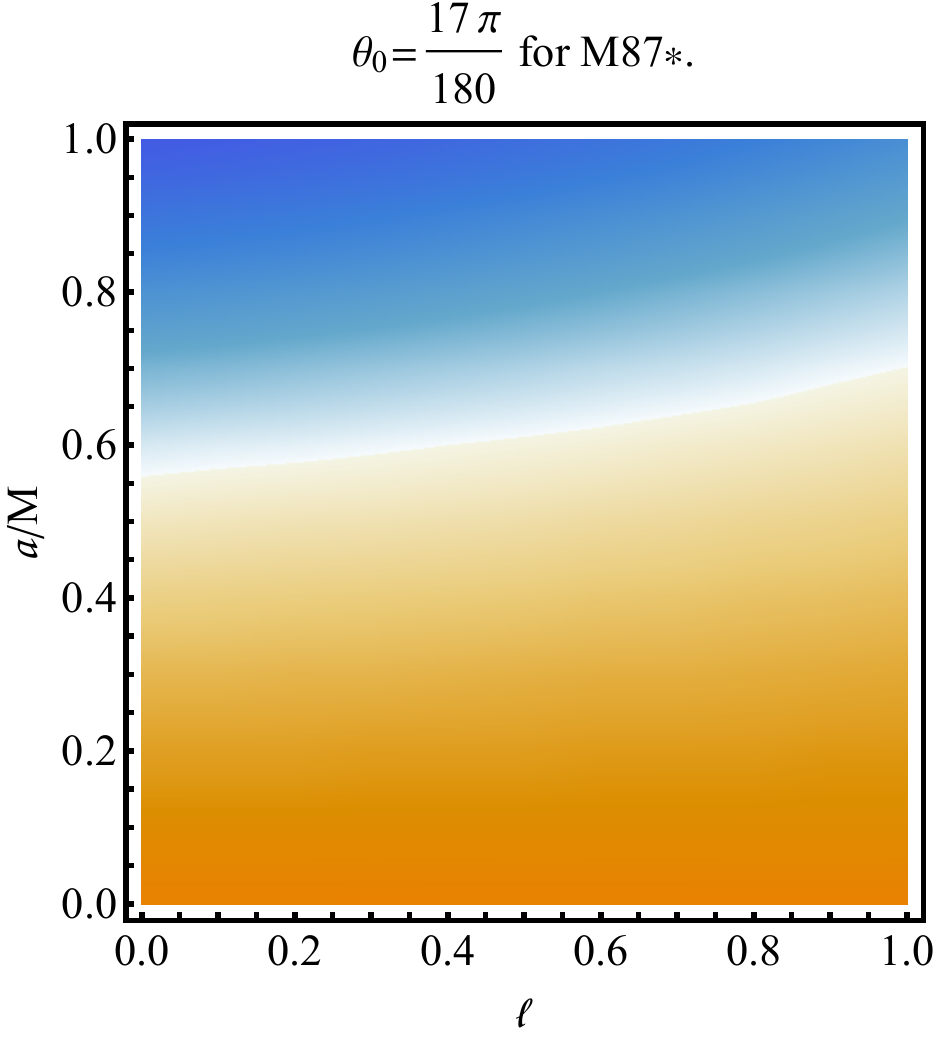}
\caption{The observable for M87* as a Ho\v{r}ava BH is expressed as a function of BH parameters, with the yellow region indicating that this parameter conforms to the observational range and the blue region suggesting its deviation from that range.}
\label{fig9}
\end{figure}
In Fig. \ref{fig9} show that parameter space of rotation Ho\v{r}ava BHs, taking M87* as an example with an observational inclination angle of $ 17\pi/180 $, we show that the yellow region indicates the parameter meets the observational range, while the blue region suggests it deviates from that range.
We find that for cases with lower spin parameters, the EHT observational data for M87* cannot impose any constraints on the Lorentz violation parameters, and the presence of Lorentz violation will allow the observational data to have a higher compatibility with the BH's spin parameters. 
For example, in Fig. \ref{fig9} with $ \ell=0 $, the spin parameter is approximately constrained to satisfy $ a/M < 0.56 $, whereas for a Ho\v{r}ava BH ($ \ell\neq0 $), it allows a spin parameter up to 0.7.
Interestingly, the spacetime described by the metric \meq{ds2} is asymptotically Minkowski at infinity, which implies that such strong Lorentz violation effects are entirely possible within the strong gravitational field around a BH.
\begin{figure}[t]
\centering
\includegraphics[width=0.8\linewidth]{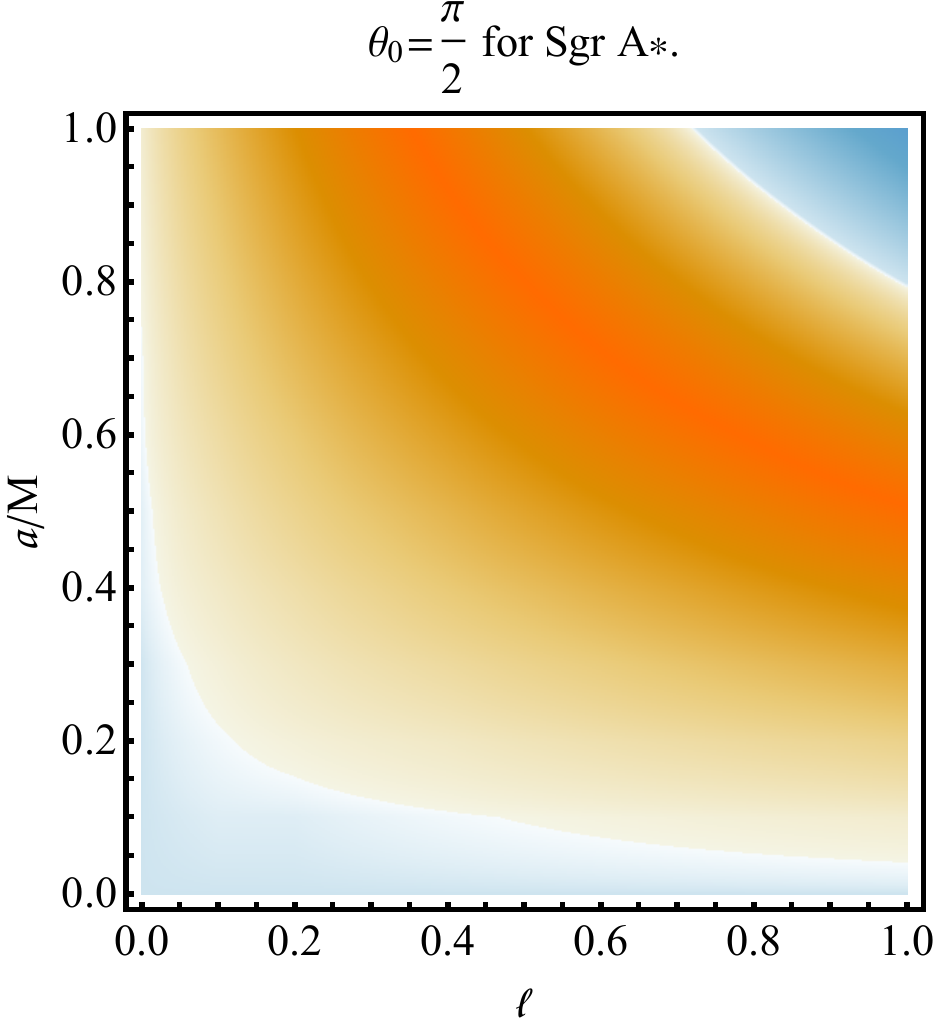}
\caption{The observable for Sgr A* as a Ho\v{r}ava BH is expressed as a function of BH parameters, with the yellow region indicating that this parameter conforms to the observational range and the blue region suggesting its deviation from that range.}
\label{fig10}
\end{figure}
Then, we can consider using the observational data of Sgr A* from the EHT to constrain the BH parameters.
For this case, an notably extreme inclination angle of $ \theta_0 = \pi/2 $ can be assumed.
Sgr A*, situated at the Galactic center, possesses a mass of $\mathcal{M} = 4.0 \times 10^6 M_\odot$ and is observed at a distance of $D_O = 8.3 kpc$ \cite{EventHorizonTelescope:2022wkp}. 
Observational data reveal that the angular diameter of the Sgr A* BH is $51.8 \pm 2.3 \mu\text{as}$ \cite{EventHorizonTelescope:2022wkp}, a measurement we will employ to impose constraints on the Lorentz-violating parameters within the Ho\v{r}ava theory. 
Fig. \ref{fig10} depicts the parameter space of the rotating Ho\v{r}ava BH, as constrained by Sgr A* observational data, under an inclination angle of $ \pi/2 $. 
In the figure, the lower-left blue region and the upper-right blue region delineate the ranges where the shadow's angular radius, as predicted by this theory, lies below the observational lower bound and above the observational upper bound, respectively, whereas the dark yellow region highlights the most likely parameter range. 
Clearly, under these conditions, the Lorentz-violating parameters retain a notably wide range of acceptable values. 
In the high-spin regime, we can establish constraints on the Lorentz-violating parameters for specific spin values. 
Within the low-spin regime, Lorentz violation may improve the alignment of observational data with the spin parameter. 
\begin{table}[t]
\renewcommand{\arraystretch}{1.25}
\centering
\setlength\tabcolsep{1.3mm}{
\begin{tabular}{ccccccccc}
\hline\hline
\multirow{2}{*}{Low-spin} & \multirow{2}{*}{ Constraints\quad}&\multirow{2}{*}{\quad High-spin} & \multirow{2}{*}{ Constraints }
\\ \\
\hline
  $a/M=0.1 $  & $ \ell>0.47 $   \quad&\quad $a/M=0.80 $  & $ \ell<0.98 $   \\
  $a/M=0.2 $  & $ \ell>0.12 $   \quad&\quad $a/M=0.85 $  & $ \ell<0.91 $    \\
  $a/M=0.3 $  & $ \ell>0.06 $   \quad&\quad $a/M=0.90 $  & $ \ell<0.84 $     \\
  $a/M=0.4 $  & $ \ell>0.03 $   \quad&\quad $a/M=0.95 $  & $ \ell<0.77 $      \\
  $a/M=0.5 $  & $ \ell>0.02 $   \quad&\quad $a/M=0.99 $  & $ \ell<0.72 $       \\
\hline\hline
\end{tabular}}
\caption{We derive parameter constraints for Sgr A* based on the Ho\v{r}ava BH metric with inclination angle $ \theta_0=\pi/2 $.}
\label{tab1}
\end{table}
Parameter constraints are detailed in Table \ref{tab1}; notably, within a spin parameter range of 0.6–0.8, existing data fail to constrain the Lorentz-violating parameters.
Should independent observations confirm that Sgr A*'s spin parameter is less than 0.6 and its inclination angle relative to Earth is $ \pi/2 $, this would constitute compelling evidence for Ho\v{r}ava gravity and the Lorentz violation it describes.

\section{Remarks}\label{Sec.5}
In this work, we have adjusted the scaling of the rotating Ho\v{r}ava BH solution to enable a more effective investigation of its shadows and associated characteristics.
We should emphasize that current observational data are insufficient to ascertain the inclination angle between Sgr A* and Earth.
This study presents the parameter space $ (\ell,a/M) $ solely for $\theta_0=\pi/2$, yet the non-separable property of photon trajectories in the rotating Ho\v{r}ava BH spacetime still demands substantial computational effort for numerical simulations.
Should one aim to explore the parameter space defined by $(\theta_0,\ell,a/M)$, the associated computational cost would become prohibitively expensive; therefore, employing neural network models \cite{Liuao2025} to lessen the dependence on CPU resources emerges as a promising future research avenue.

~~\\~~
\acknowledgments
This work was supported by the National Natural Science Foundation of China under Grants No. 12122504, No. 12375046, No. 12035005, No. 12205243, No. 12375053, by the innovative research group of Hunan Province under Grant No. 2024JJ1006, by the Natural Science Foundation of Hunan Province under grant No. 2023JJ30384, by the Hunan provincial major sci-tech program under grant No.2023zk1010, by the Sichuan Science and Technology Program under Grant No. 2023NSFSC1347, and by the Doctoral Research Initiation Project of China West Normal University under Grant No. 21E028.

%

\appendix
\begin{widetext}
\section{Trajectory equations}\label{AppendixA}

In this Appendix, we show the null geodesic equations governing the photon trajectories in the gravitational field of rotating BHs within the low-energy sector of Ho\v{r}ava gravity, which are given by the following equations.
\begin{align}
&\begin{aligned}
\dot{t}=\frac{\Sigma^2\mathcal{E}}{\rho^2\Delta}-\frac{2MraL_z(\ell+1)}{\rho^2\Delta},
\end{aligned}\\[0.5ex]
&\begin{aligned}
\ddot{r}=&\left(\frac{\pp_r\Delta}{2\Delta}-\frac{\pp_r\rho^2}{2\rho^2}\right)\dot{r}^2-\frac{\pp_\theta\rho^2}{\rho^2}\dot{r}\dot{\theta}+\frac{\Delta\pp_r\rho^2}{2\rho^2}\dot{\theta}^2
+\frac{M\Delta\left(\rho^2-r\pp_r\rho^2\right)}{\rho^6}\left(\dot{t}-a\sin^2\theta~\dot{\varphi}\right)^2
+\frac{r\Delta\sin^2\theta}{\rho^2}\dot{\varphi}^2\\
&-\frac{2Ma\ell\Delta\sin^2\theta}{\rho^6}\left(\rho^2-r\pp_r\rho^2\right)\dot{r}\dot{\varphi}
-\frac{2M^2r^2a^2\ell(\ell+2)\Delta\sin^2\theta}{\rho^4\Sigma^2}\left(\frac{\pp_r\rho^2}{\rho^2}+\frac{\pp_r\Sigma^2}{\Sigma^2}-\frac{2}{r}\right)\dot{t}^2,
\end{aligned}\\[0.5ex]
&\begin{aligned}
\ddot{\theta}=&\frac{\pp_\theta\rho^2}{2\rho^2}\left(\frac{\dot{r}^2}{\Delta}-\dot{\theta}^2\right)-\frac{\pp_r\rho^2}{\rho^2}\dot{r}\dot{\theta}
-\frac{\pp_\theta\rho^2}{\rho^2}\left(\frac{Mr}{\rho^4}\dot{t}^2+\frac{r^2+a^2}{a^2}\dot{\varphi}^2\right)
-\frac{\sin^2\theta Mr\pp_\theta\rho^2}{2\rho^4}\left(\frac{\tan\theta\pp_\theta\rho^2}{\rho^2}+4\right)\dot{\varphi}^2\\
&+\frac{2Mra(\ell+1)\sin^2\theta}{\rho^4}\left[\left(\frac{\pp_\theta\rho^2}{\rho^2}-2\cot\theta\right)\dot{t}\dot{\varphi}
-\frac{Mra\ell(\ell+2)}{\rho^4\Sigma^2(\ell+1)}\left(\frac{\pp_\theta\Sigma^2}{\Sigma^2}+\frac{\pp_\theta\rho^2}{\rho^2}-2\cot\theta\right)\dot{t}^2\right],
\end{aligned}\\[1.5ex]
&\begin{aligned}
\dot{\varphi}=\frac{2Mra\mathcal{E}-a^2L_z}{\rho^2\Delta}+\frac{L_z}{\rho^2\sin^2\theta}+\frac{2Mra\ell}{\rho^2\Delta}
\left[\mathcal{E}-\frac{2MraL_z(\ell+2)}{\Sigma^2}\right],
\end{aligned}
\end{align}
with the photon Hamiltonian constraint condition
\begin{align}\label{HHHHH}
\mathcal{H}=&
\frac{\left(L_z\csc\theta-a\mathcal{E}\sin\theta\right)^2}{2\rho^2}-\frac{[(r^2+a^2)\mathcal{E}-aL_z]^2}{2\rho^2\Delta}+\frac{\rho^2\dot{r}^2}{2\Delta}+\frac{\rho^2 \dot{\theta}^2}{2}
+\frac{2L_zMra\ell\mathcal{E}}{\rho^2\Delta}-\frac{2L_z^2M^2r^2a^2\ell(\ell+2)}{\rho^2\Delta\Sigma^2}=0.
\end{align}

\section{Numerical backward ray-tracing method}\label{AppendixB}
In this appendix, we introduce the details of the numerical backward ray-tracing method.
Given that the spacetime of a Kerr-like rotating compact object is asymptotically flat, the observer's orthonormal basis $ \{ e_{\hat{t}},e_{\hat{r}},e_{\hat{\theta}},e_{\hat{\varphi}}  \} $ can be expressed in terms of the coordinate basis $ \{ \pp_t,  \pp_r,  \pp_\theta,  \pp_\varphi  \} $ as
\begin{equation}
e_{\hat{\mu}}=e^{~\nu}_{\hat{\mu}}\pp_\nu,
\end{equation}
where the transformation matrix $ e^{~\nu}_{\hat{\mu}} $ satisfies the condition
\begin{equation}\label{geta}
g_{\mu\nu}e^{~\mu}_{\hat{\alpha}}e^{~\nu}_{\hat{\beta}}=\eta_{\hat{\alpha}\hat{\beta}},
\end{equation}
and the $ \eta_{\hat{\alpha}\hat{\beta}} $ represents the Minkowski metric.
For the rotating spacetime in the low-energy sector of Ho\v{r}ava gravity, it is convenient to choose a decomposition based on a reference frame having zero axial angular momentum at spatial infinity 
\begin{equation}\label{xishu}
e^{~\nu}_{\hat{\mu}}=\left( \begin{array}{cccc}
\zeta &0  &0  &\gamma  \\
0 &A^r  &0  &0  \\
0 &0  &A^\theta  & 0 \\
0 &0  &0  &A^\varphi 
\end{array}  \right),
\end{equation}
where $ \zeta $, $ \gamma $, $ A^r $, $ A^\theta $ and $ A^\varphi $ are real coefficients.
Based on the Minkowski normalization,  
\begin{equation}
e_{\hat{\mu}} e^{\hat{\nu}} = \delta^{~\hat{\nu}}_{\hat{\mu}},
\end{equation}  
we solve this relationship \meq{geta} to obtain the real coefficients of the transformation matrix \( e^{~\nu}_{\hat{\mu}} \):  
\begin{equation}
A^r = \frac{1}{\sqrt{g_{rr}}}, ~~~~~~~~ A^\theta = \frac{1}{\sqrt{g_{\theta\theta}}}, ~~~~~~~~ A^\varphi = \frac{1}{\sqrt{g_{\varphi\varphi}}},~~~~~~~~
\zeta = \sqrt{\frac{g_{\varphi\varphi}}{g^2_{t\varphi} - g_{tt} g_{\varphi\varphi}}}, ~~~~~~
\gamma = -\frac{g_{t\varphi}}{g_{\varphi\varphi}} \sqrt{\frac{g_{\varphi\varphi}}{g^2_{t\varphi} - g_{tt} g_{\varphi\varphi}}}.
\end{equation}  
We then derive the orthogonal tetrad form of \( e_{\hat{\mu}} \) as follows:  
\begin{eqnarray}\label{zjbj}
e_0&=& e_{\hat{t}} = \left. \sqrt{\frac{g_{\varphi\varphi}}{g_{t\varphi}^2 - g_{tt} g_{\varphi\varphi}}} \left( \partial_t - \frac{g_{t\varphi}}{g_{\varphi\varphi}} \partial_\varphi \right) \right|_{(r_0, \theta_0)},\\
e_1&=& -e_{\hat{r}} = -\left. \sqrt{\frac{1}{g_{rr}}} \partial_r \right|_{(r_0, \theta_0)},\\
e_2&=& e_{\hat{\theta}} = \left. \frac{1}{\sqrt{g_{\theta\theta}}} \partial_\theta \right|_{(r_0, \theta_0)},\\
e_3&=& -e_{\hat{\varphi}} = -\left. \frac{1}{\sqrt{g_{\varphi\varphi}}} \partial_\varphi \right|_{(r_0, \theta_0)}.
\end{eqnarray}
Here, $ g_{\mu\nu} $ represents the BH metric components in the rotating spacetime of the low-energy sector of Hořava gravity. 
It should be emphasized that this tetrad is not unique; an appropriate tetrad can be selected based on specific requirements.
In the chosen tetrad \meq{zjbj}, $e_0$ represents the four-velocity of the observer, while $e_3$ points in the spatial direction toward the BH's center. 
Additionally, the linear combinations $ e_0\pm e_3 $ are tangential to the principal null congruences of the metric.

\begin{figure}
\centering
\includegraphics[width=0.45\linewidth]{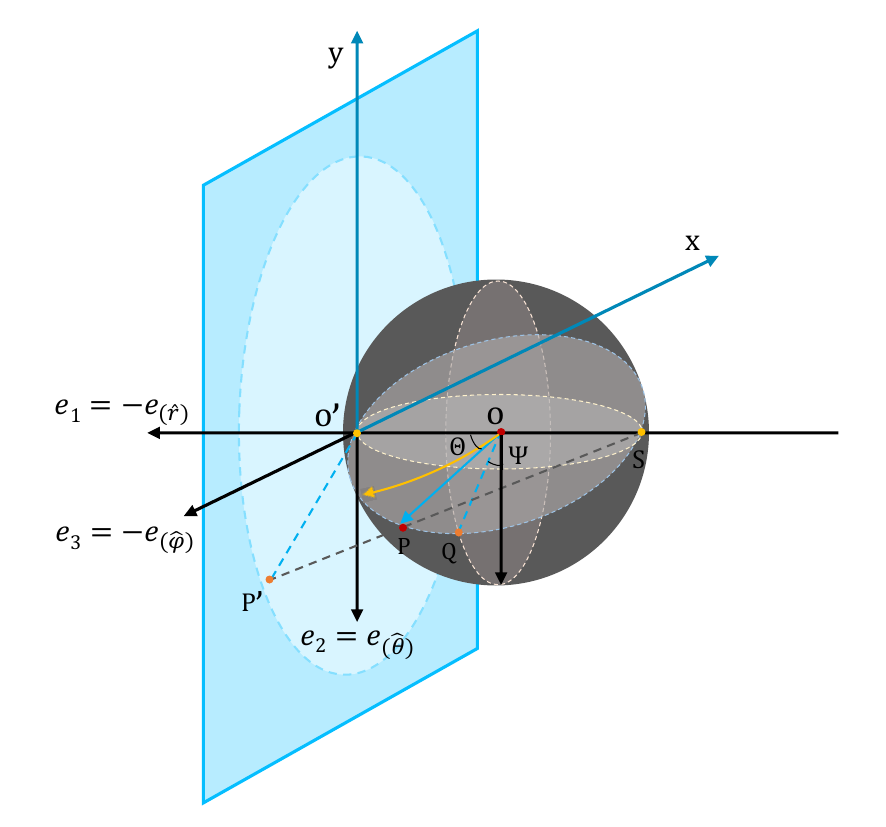}
\caption{In the observer's frame, the celestial coordinates $ \Theta $ and $ \Psi $ are introduced to identify each light ray. 
Using stereographic projection, a useful map is then created that translates the celestial sphere onto the camera screen.}
\label{fig1}
\end{figure}
To describe the position of the photon observed, a celestial coordinate system is introduced.
As shown in the Fig. \ref{fig1}, the observer is located at point $O$, and the light propagates along the null geodesic. 
For practical calculations, the light can be considered as originating from the observer due to the reversibility of optical paths.
In three-dimensional space, the tangent vector of the null geodesic at point $O$ is denoted as $\overrightarrow{OP}$.
A sphere is constructed with $O$ as the center and $|\overrightarrow{OP}|$ as the radius. 
The diameter $O'S$ is parallel to $e_1$ and lies on the equatorial plane of the sphere. 
The angle between the line $OO'$ and $OP$ is defined as the first celestial coordinate, $\Theta$. 
Furthermore, the plane formed by $OO'$ and $OP$ intersects the sphere to form a great circle passing through points $P$, $S$, and $O'$. 
This great circle intersects another great circle, perpendicular to $e_1$, at two points. 
Among these, the intersection point closer to $P$ is denoted as $Q$. 
The angle between the vector $\overrightarrow{OQ}$ and $e_2$ is defined as the second celestial coordinate, $\Psi$. Thus, the complete celestial coordinates of the photon are given as $(\Theta, \Psi)$.

On the other hand, for each light ray $ \lambda(s) $,  described in coordinates as $ t(s), r(s),\vartheta(s),\varphi(s) $, has a general tangent vector given by
\begin{equation}\label{lambda1}
\dot{\lambda}=\dot{t}\pp_t+\dot{r}\pp_r+\dot{\vartheta}\pp_\vartheta+\dot{\varphi}\pp_\varphi.
\end{equation}
In the observer's reference frame, the tangent vector of the null geodesic can be expressed in terms of the orthonormal tetrad as
\begin{equation}
\dot{\lambda}=|\overrightarrow{OP}|(- \chi e_0+\sin\Theta\cos\Psi e_1+\sin\Theta\sin\Psi e_2+\cos\Theta e_3),
\end{equation}
where $\chi$ is a scalar factor, and $ |\overrightarrow{OP}| $ represents the magnitude of the tangent vector of the null geodesic at a point $ O $ in three-dimensional space.
Using this representation, we can derive a stereographic projection that maps points from the celestial sphere onto a plane. 
The projection is described by the following equations:
\begin{equation}\label{B4}
x_{P'}=-2|\overrightarrow{OP}|\tan\frac{\Theta}{2}\sin\Psi,\quad\quad
y_{P'}=-2|\overrightarrow{OP}|\tan\frac{\Theta}{2}\cos\Psi,
\end{equation}
where $ P' $ is the projection of a point $ P $ on the celestial coordinate system onto the plane of the Cartesian coordinate system.
To visualize this projection, one effective method is to employ a pinhole camera for perspective projection \cite{Hu:2020usx}. 
This model, which aligns with practical imaging principles and is simple, is commonly referred to as the fisheye camera model, but it is limited by a relatively narrow field of view.
For the plane, we can establish a standard Cartesian coordinate system with origin $ O' $, as illustrated in Fig. \ref{fig2}.
The field of view angle $ \alpha $, defined as the angle between $ O $ and the projection screen, is simplified by assuming that the angular extents in both the $ yO'z $ and  $ xO'z $ planes are equal.
\begin{figure}[h]
\centering
\includegraphics[width=0.5\linewidth]{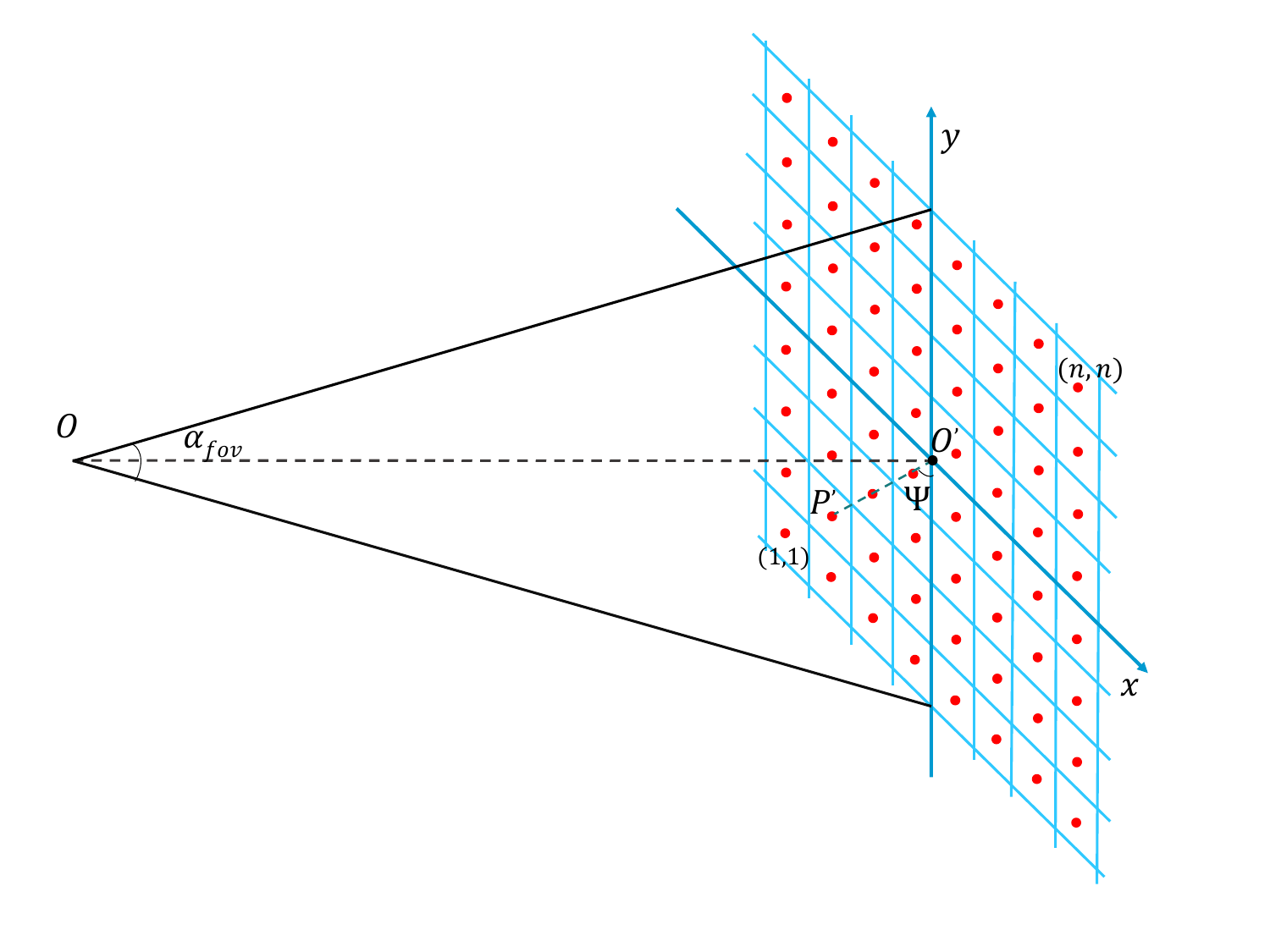}
\caption{
Illustration of the pixels. 
A standard Cartesian coordinates with the origin at $ O' $ are defined in the field of view.}
\label{fig2}
\end{figure}

Then, to determine the dimensions of the square screen, we calculate its side length $ L $, which is given by
\begin{equation}
L=2|\overrightarrow{OP}|\tan\frac{\alpha_\text{~fov}}{2}.
\end{equation}
The screen consists of $ n\times n $ pixels, each occupying a square with a side length
\begin{equation}
l=\frac{2|\overrightarrow{OP}|}{n}\tan\frac{\alpha_\text{~fov}}{2}.
\end{equation}
Pixels are labeled by indices $ (i,j) $, where $ i $ and $ j $ range from $ 1 $ to $ n $.
The bottom-left pixel is labeled as $ (1,1) $, while the top-right pixel is $ (n,n) $.
The Cartesian coordinates of the center of a pixel are given by
\begin{equation}\label{B7}
x_{P'}=l \left(i-\frac{n+1}{2}\right), ~~~~~~ y_{P'}=l\left(j-\frac{n+1}{2}\right).
\end{equation}
By comparing the coordinates in equations \meq{B4} and \meq{B7}, we derive a relationship between the pixel indices $ (i,j) $ and the angular coordinates $ (\Theta,\Psi) $:
\begin{equation}
\tan\Psi=\frac{j-(n+1)/2}{i-(n+1)/2},\quad\quad
\tan\frac{\Theta}{2}=\tan\frac{\alpha_\text{~fov}}{2}\frac{\sqrt{[i-(n+1)/2]^2+[j-(n+1)/2]^2}}{n}.
\end{equation}

\section{Shadows and Lensing rings}\label{AppendixC}
In this Appendix, we present the complete information regarding light rays around the BH.
Before presenting these results, it should be noted that an extended light source was used to illuminate the system. 
\begin{figure}[h]
\centering
\includegraphics[width=0.45\linewidth]{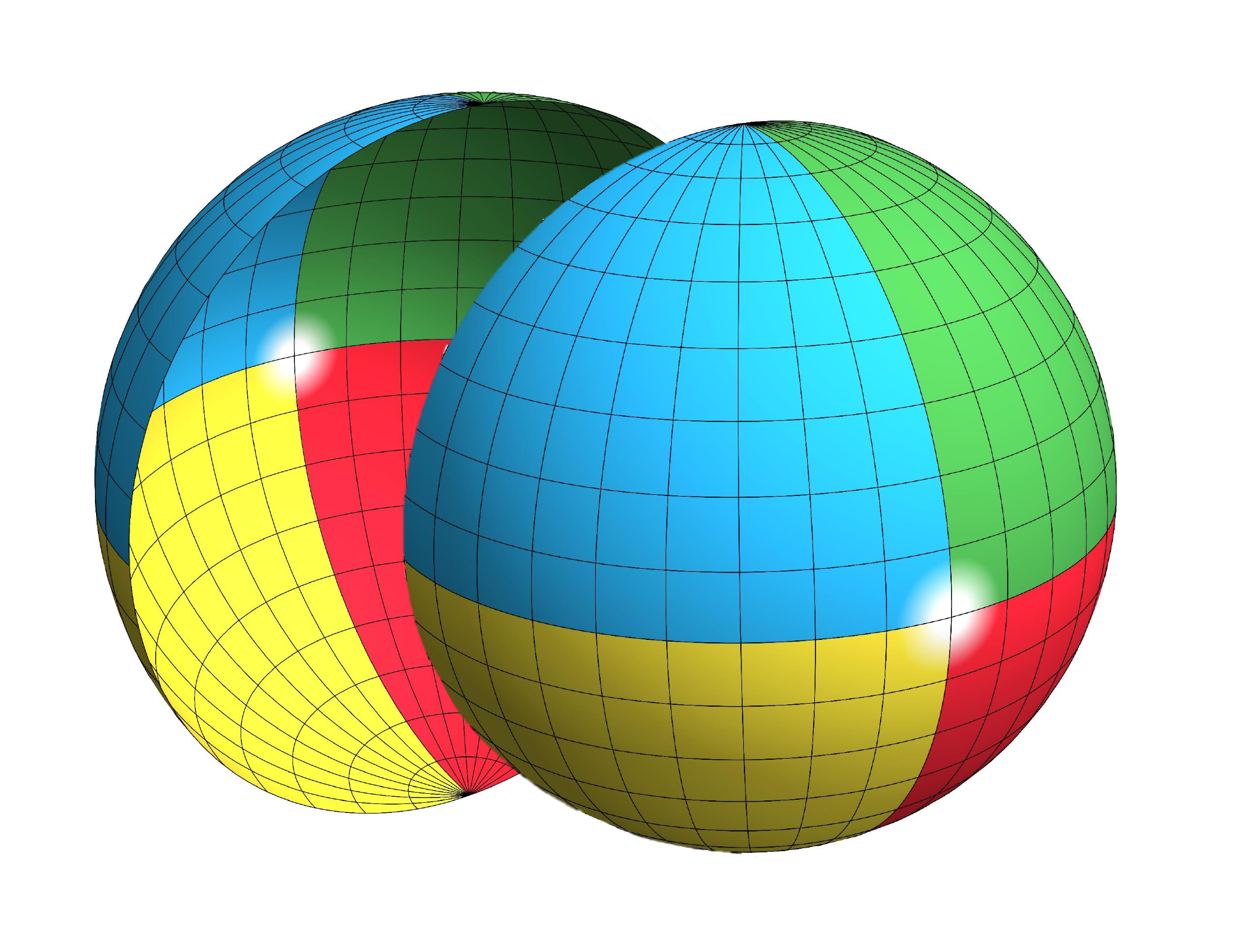}
\caption{Illustration of our spherical light source at infinity.}
\label{fig6}
\end{figure}
Fig. \ref{fig6} shows a cross-sectional view of the "large sphere" encompassing both the BH and the observer, divided into a grid with latitude and longitude lines spaced at intervals of $\pi/18$.
The BH image was generated by assigning colors to each grid segment, with the extended source divided into four distinct colors. The color of each pixel was determined by tracing photon paths from the source. 
The origin point, marked in white, appears as a lensing ring due to gravitational effects. 
Dark regions indicate photons absorbed by the BH.

Figs. \ref{fig4} and \ref{fig5} respectively illustrate the shadow contour and the lensing ring for positive and negative Lorentz violation parameters.
Compared to Fig. \ref{fig1n} in the main text, these figures reveal more physical information by considering a wider range of parameter values, enabling a more intuitive understanding of the impact of these parameter values on the BH image.
A positive Lorentz violation slightly enhances the frame-dragging effect in a rotating BH, while its impact on the size of the lensing ring remains negligible.
However, as the negative Lorentz violation intensity increases, it significantly weakens the frame-dragging effect inside the lensing ring, nearly eliminating it entirely.
This is not merely a cancellation of the BH's rotational effects, as it may seem for smaller spin parameters; rather, it fundamentally alters the BH's shadow characteristics.
Furthermore, the shadow can appear flattened, resembling that of a rotating BH but lacking frame-dragging inside the lensing ring, as illustrated in Figs. \ref{fig5}(l) and \ref{fig5}(p).
\begin{figure}[h]
\centering
\includegraphics[width=0.22\linewidth]{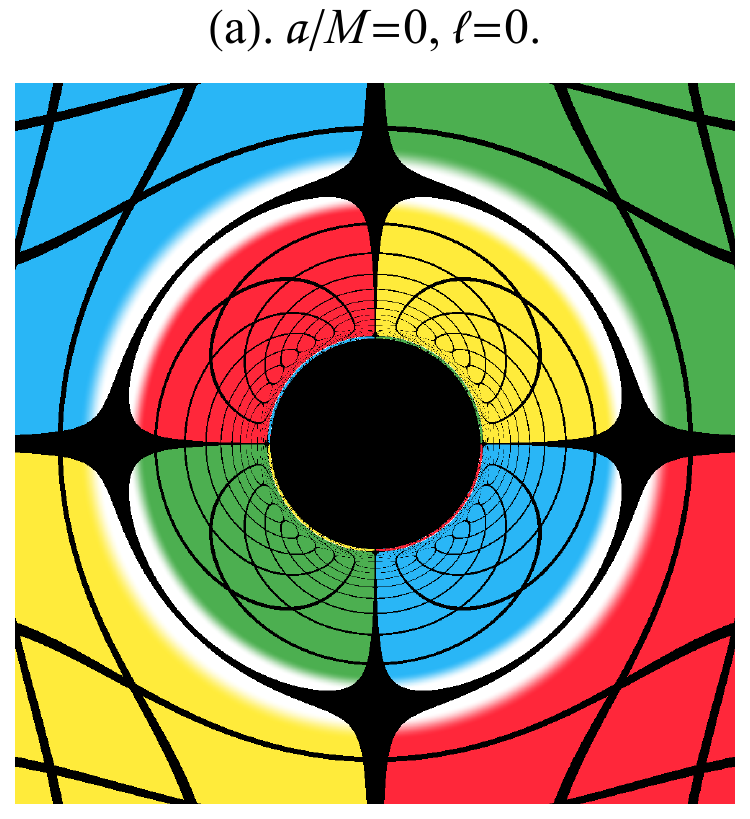}
\includegraphics[width=0.22\linewidth]{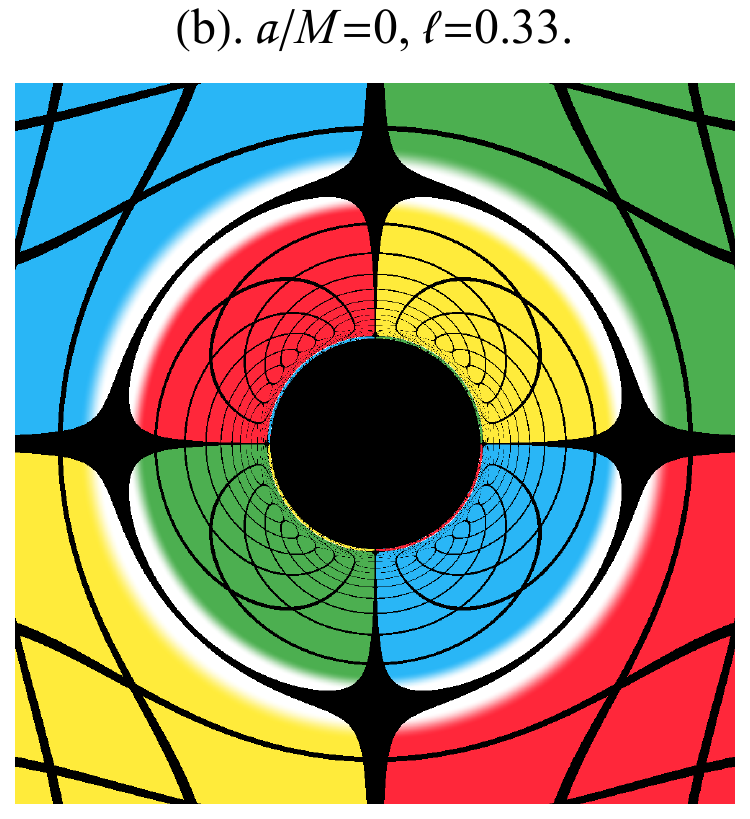}
\includegraphics[width=0.22\linewidth]{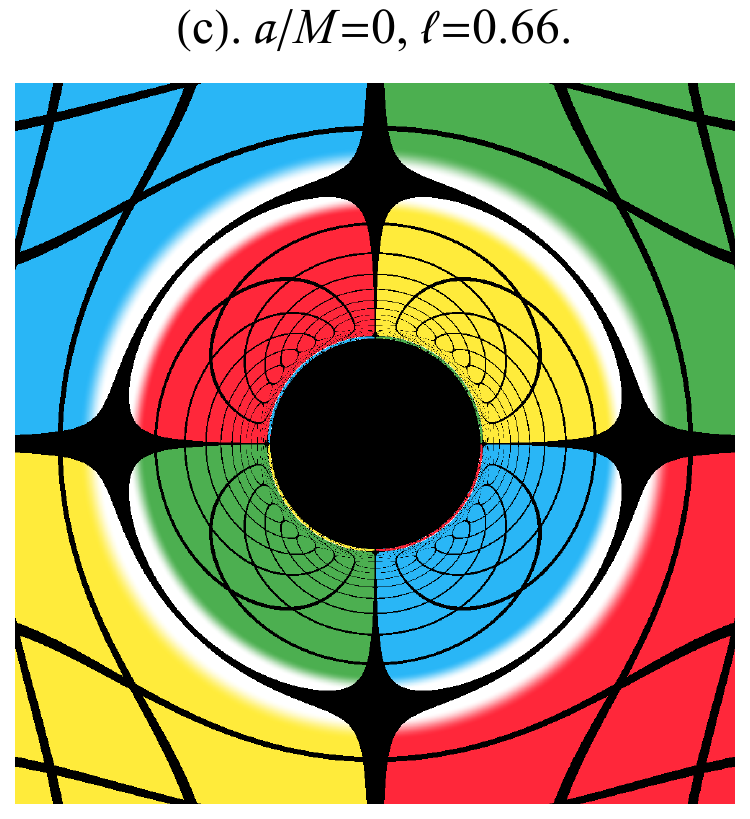}
\includegraphics[width=0.22\linewidth]{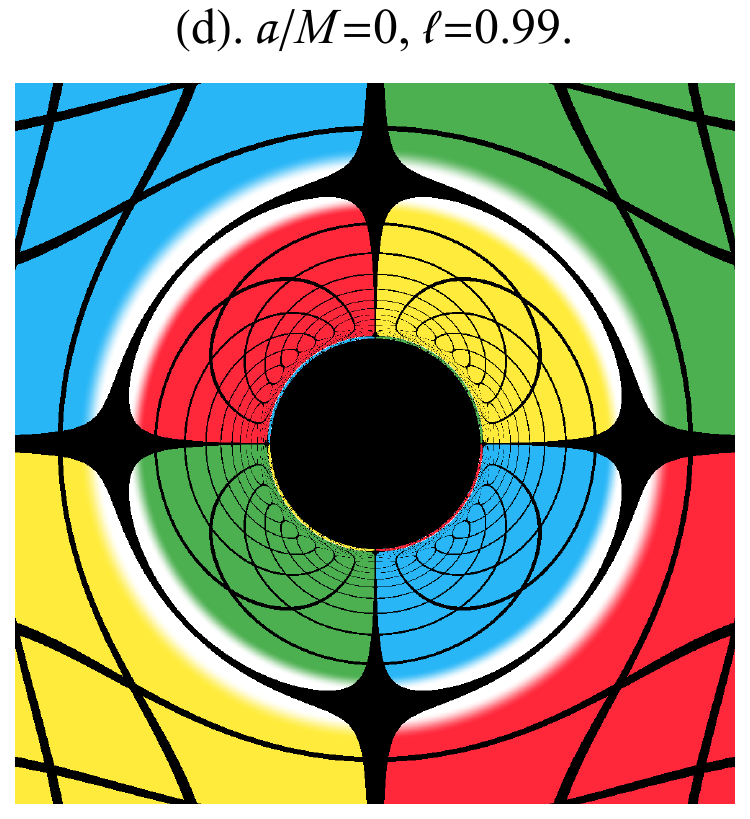}
\includegraphics[width=0.22\linewidth]{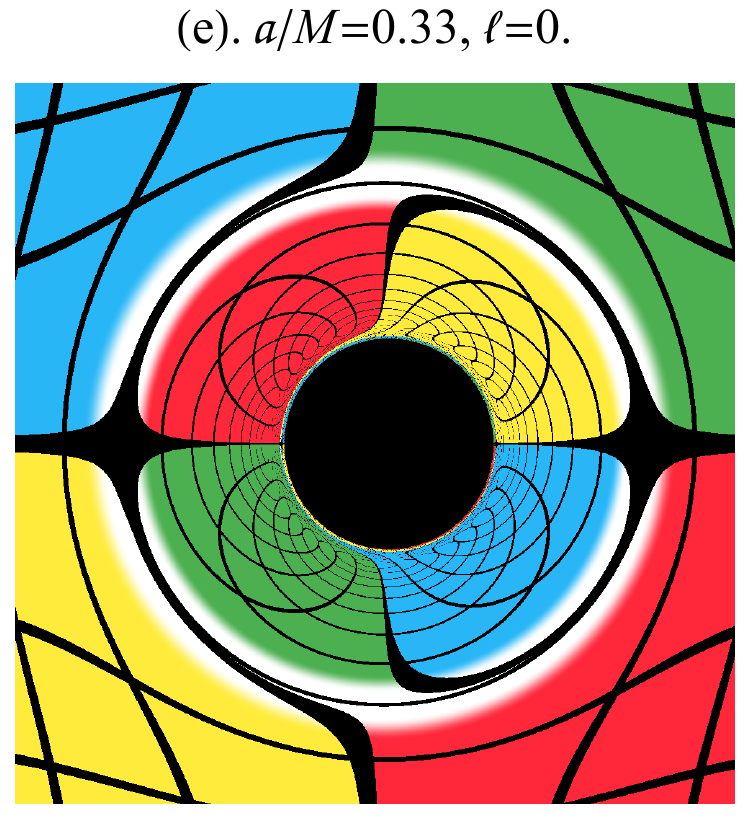}
\includegraphics[width=0.22\linewidth]{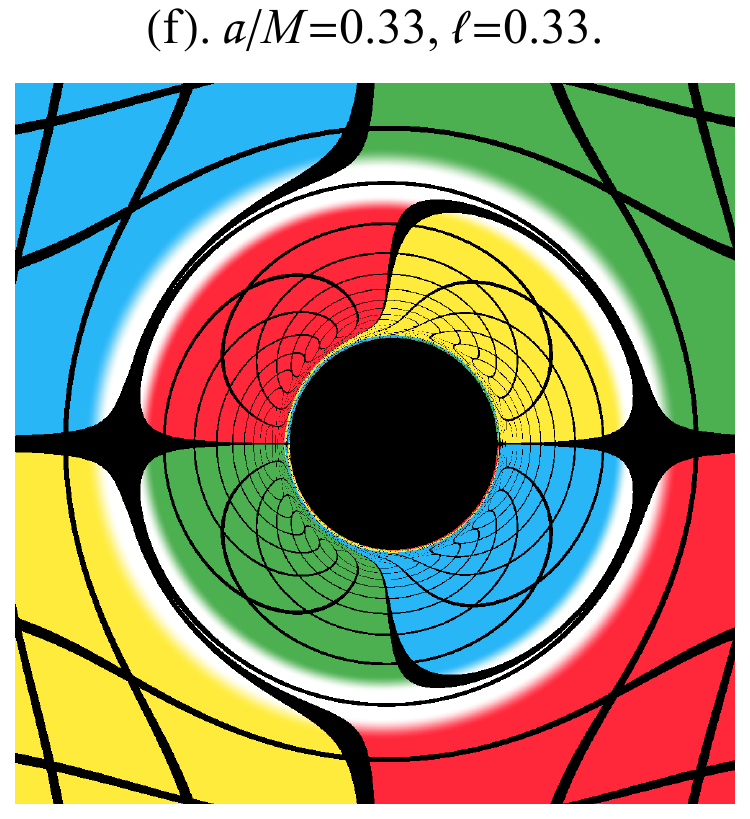}
\includegraphics[width=0.22\linewidth]{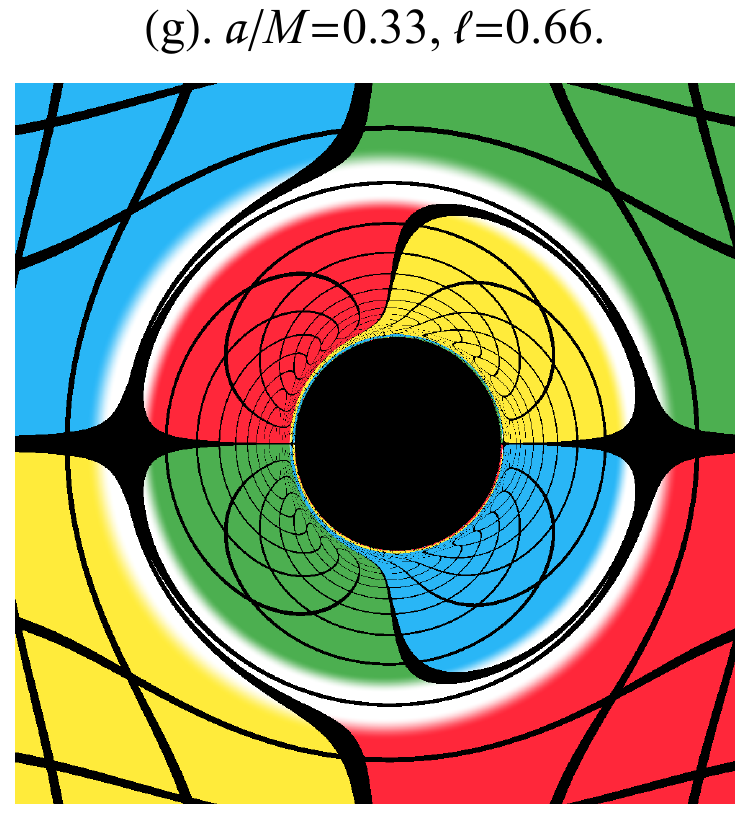}
\includegraphics[width=0.22\linewidth]{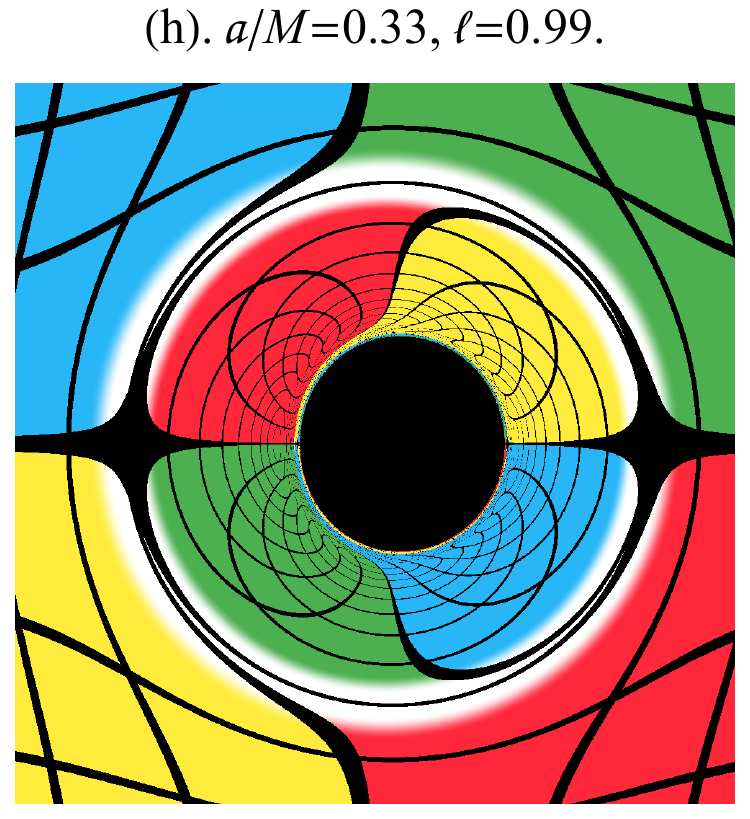}
\includegraphics[width=0.22\linewidth]{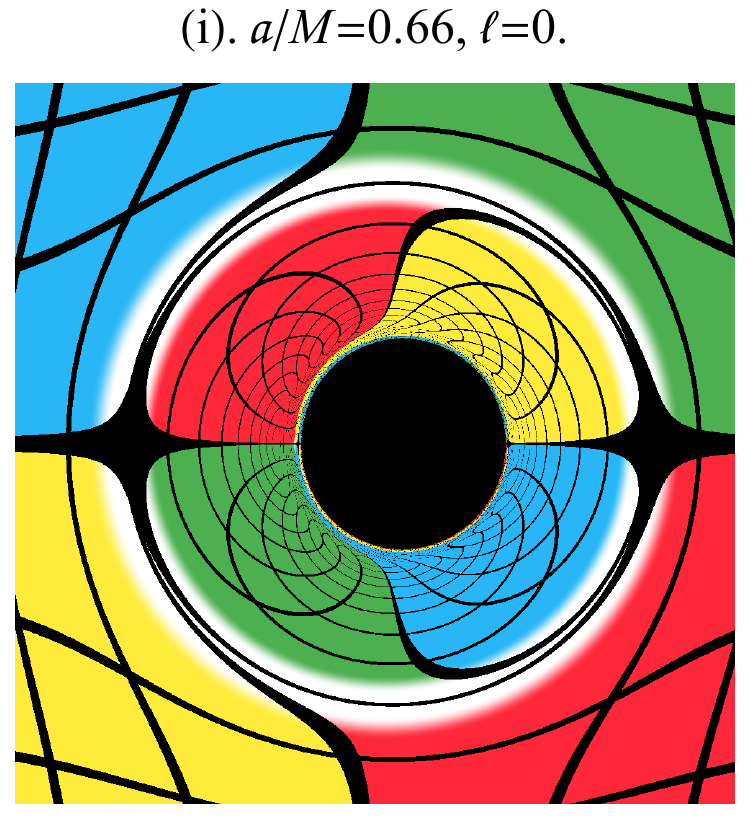}
\includegraphics[width=0.22\linewidth]{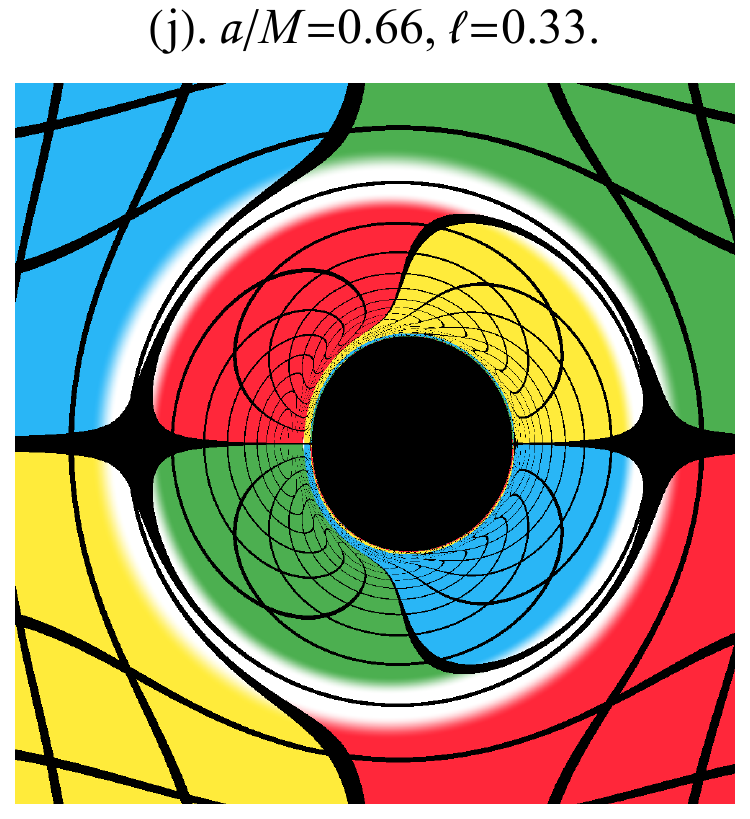}
\includegraphics[width=0.22\linewidth]{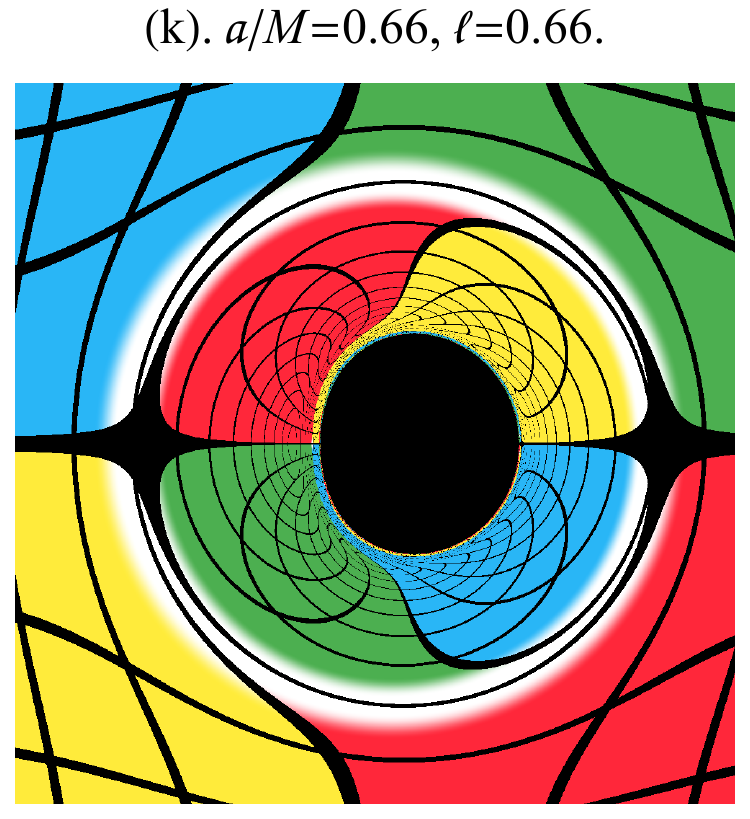}
\includegraphics[width=0.22\linewidth]{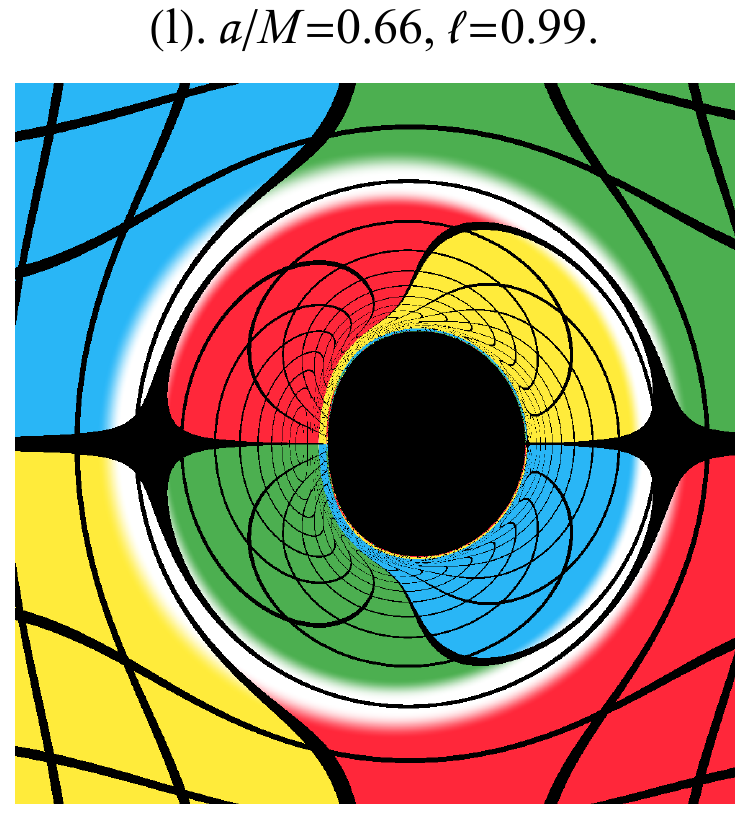}
\includegraphics[width=0.22\linewidth]{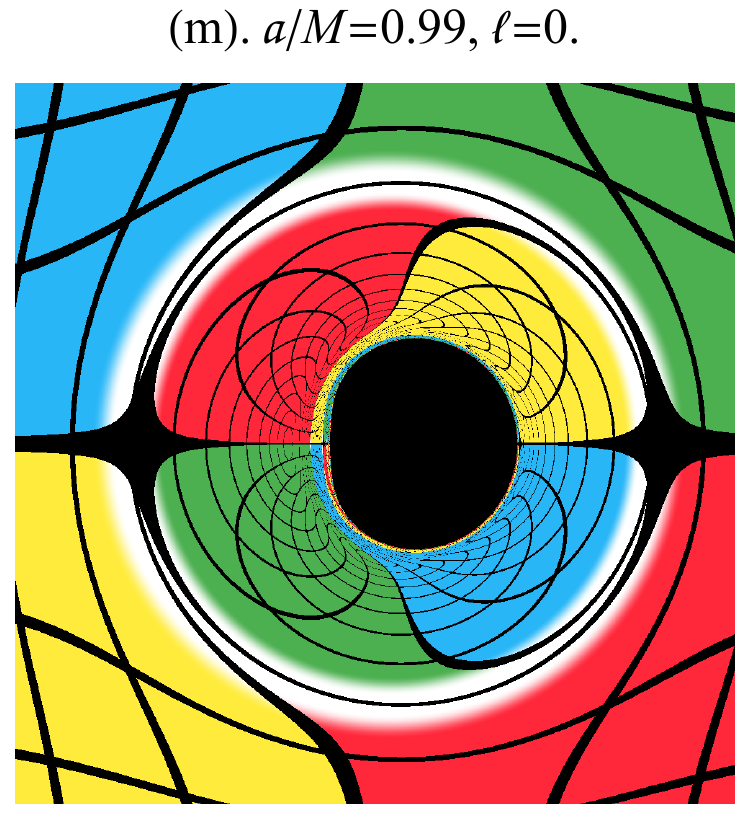}
\includegraphics[width=0.22\linewidth]{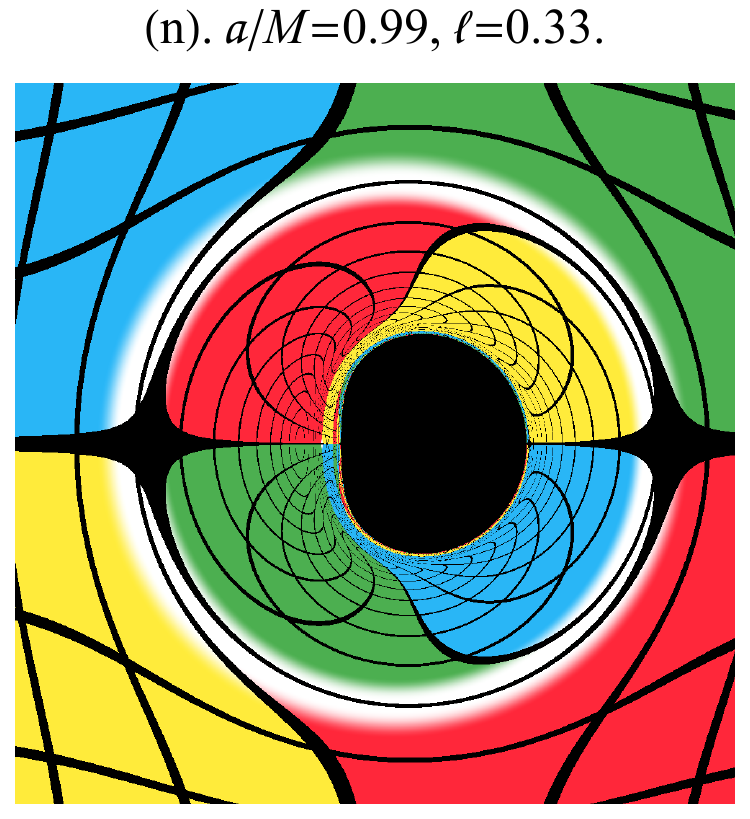}
\includegraphics[width=0.22\linewidth]{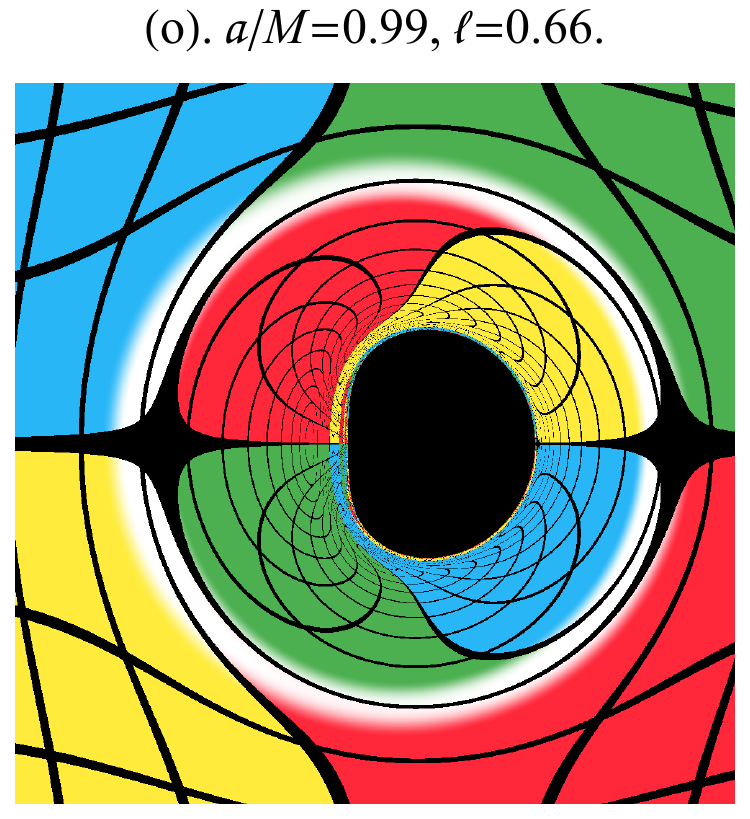}
\includegraphics[width=0.22\linewidth]{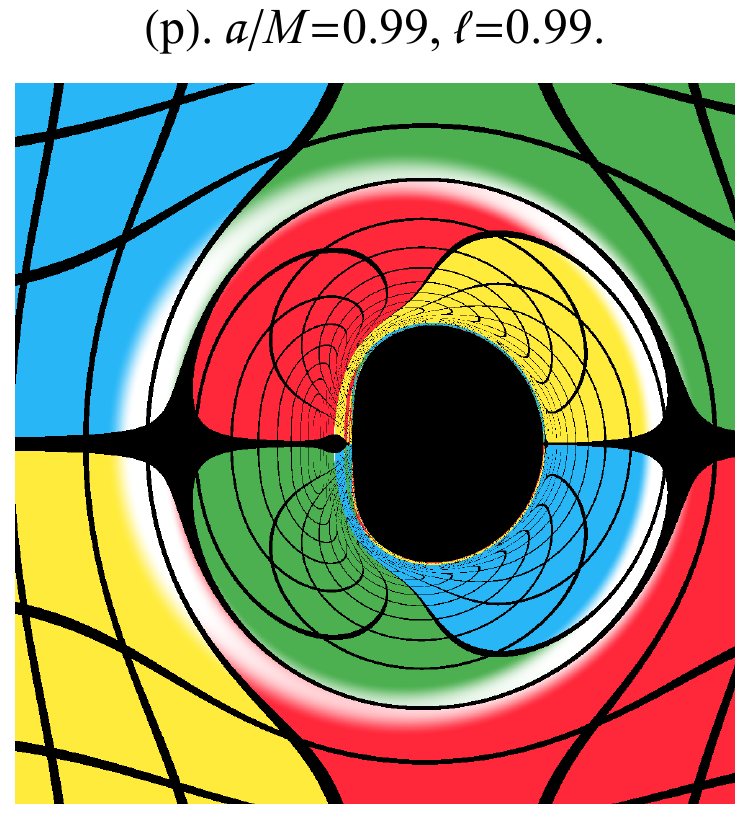}
\caption{Shadows cast by rotating BHs with a positive Lorentz violation parameter $ \ell $, as seen by an observer at $ \theta_0=\pi/2 $.}
\label{fig4}
\end{figure}
\begin{figure}[h]
\centering
\includegraphics[width=0.22\linewidth]{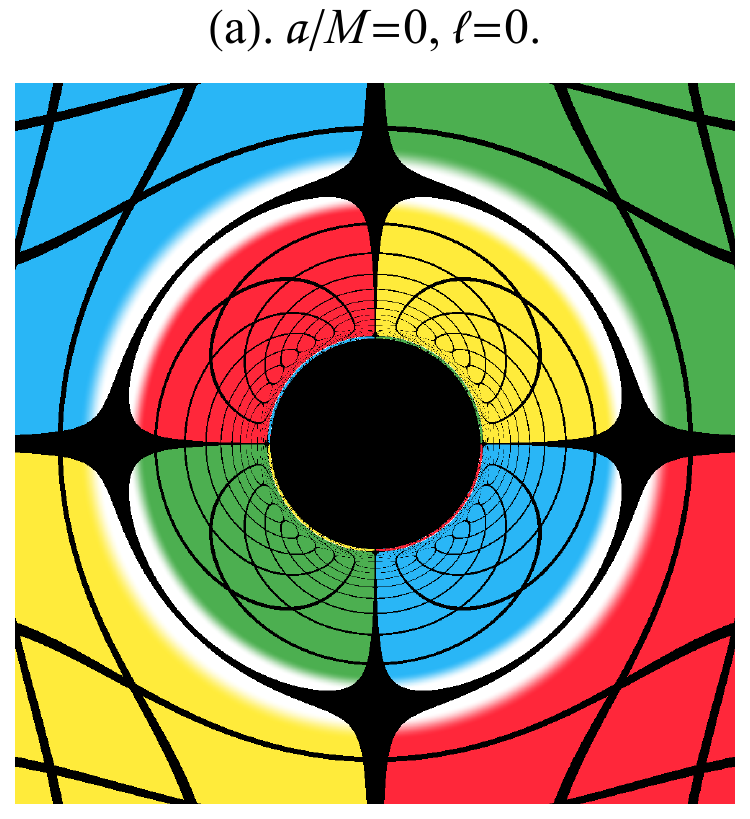}
\includegraphics[width=0.22\linewidth]{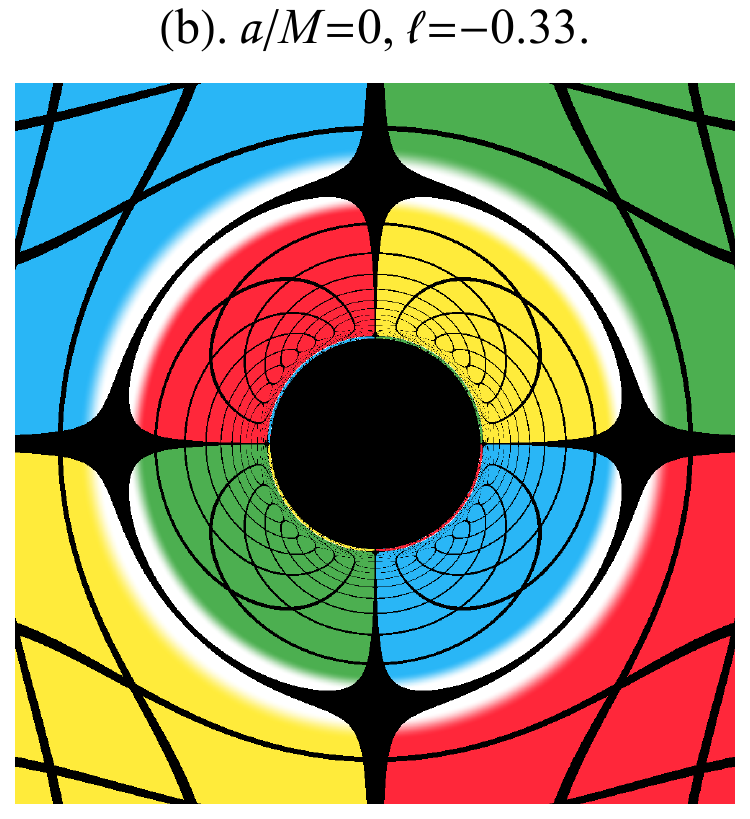}
\includegraphics[width=0.22\linewidth]{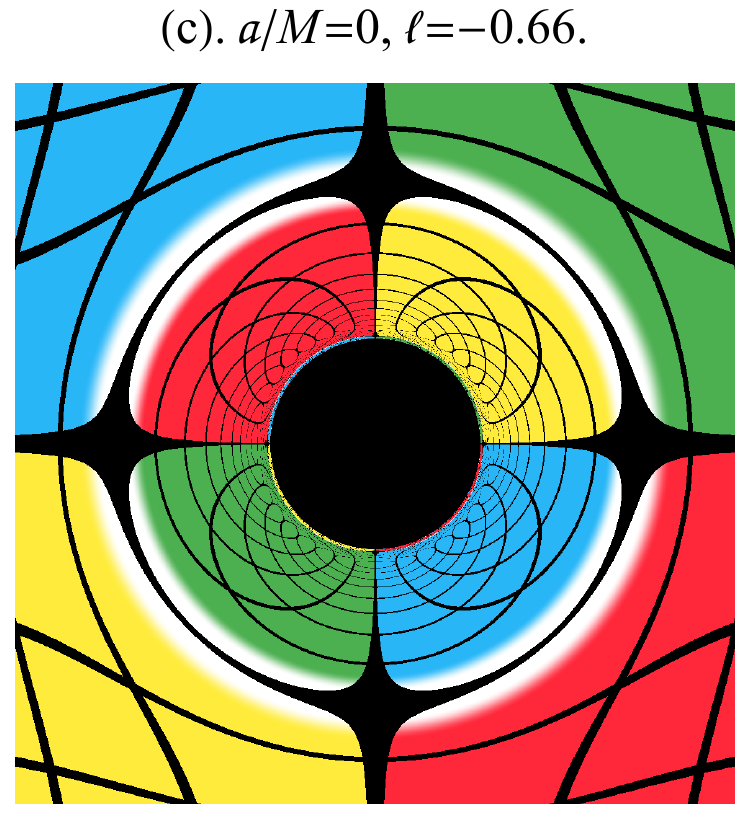}
\includegraphics[width=0.22\linewidth]{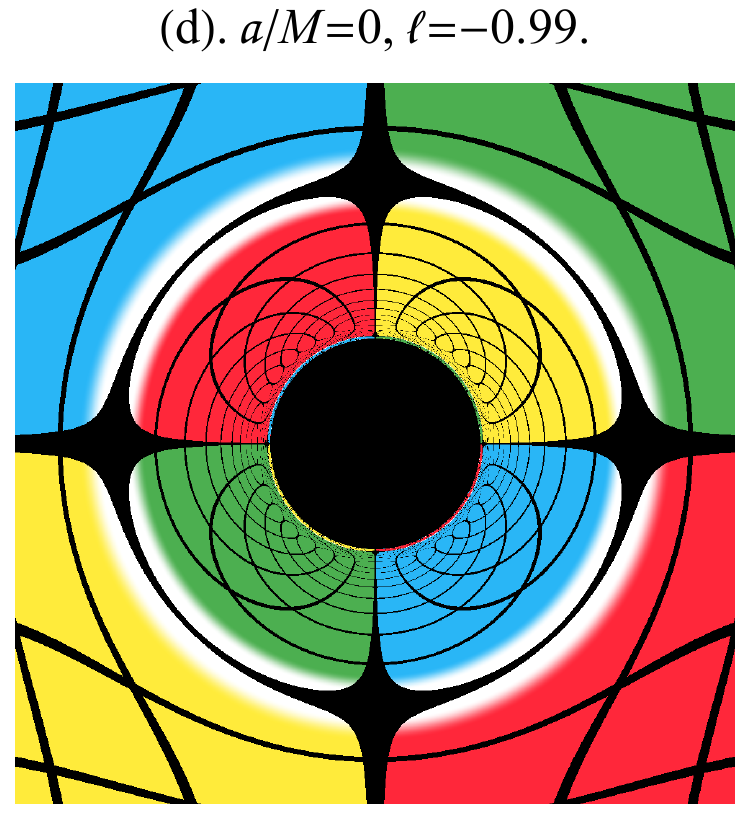}
\includegraphics[width=0.22\linewidth]{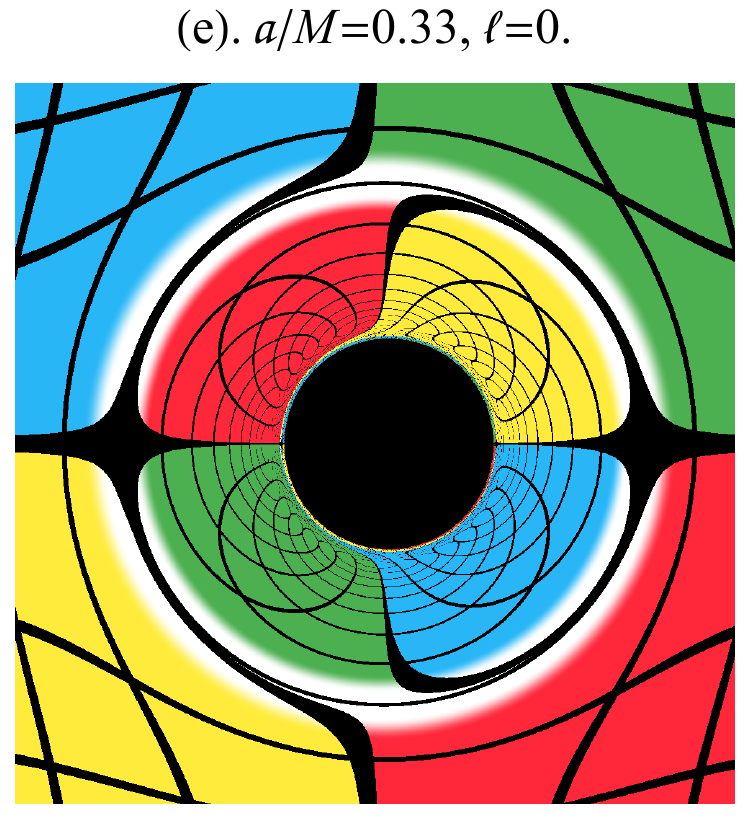}
\includegraphics[width=0.22\linewidth]{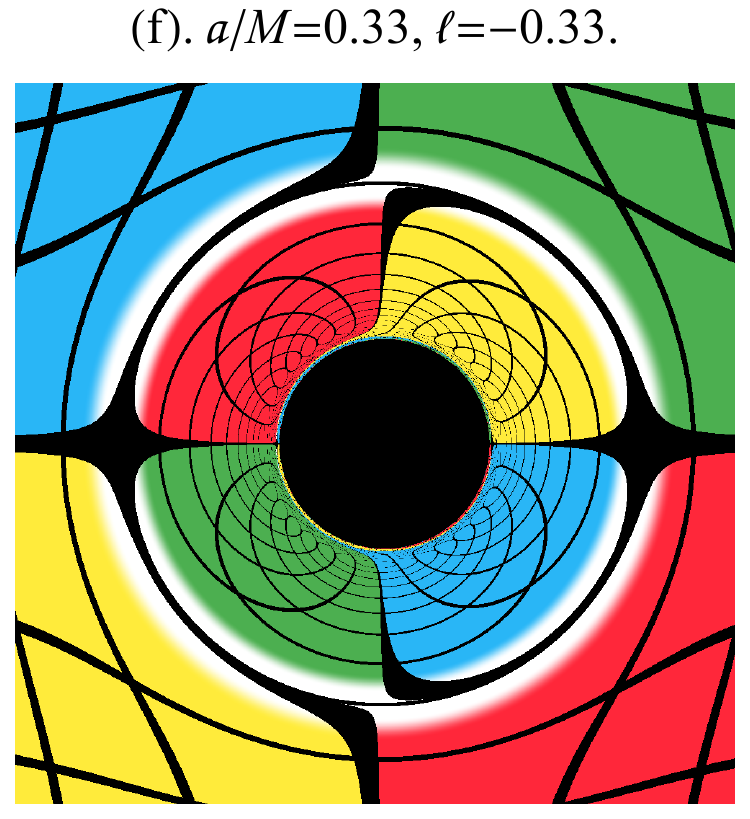}
\includegraphics[width=0.22\linewidth]{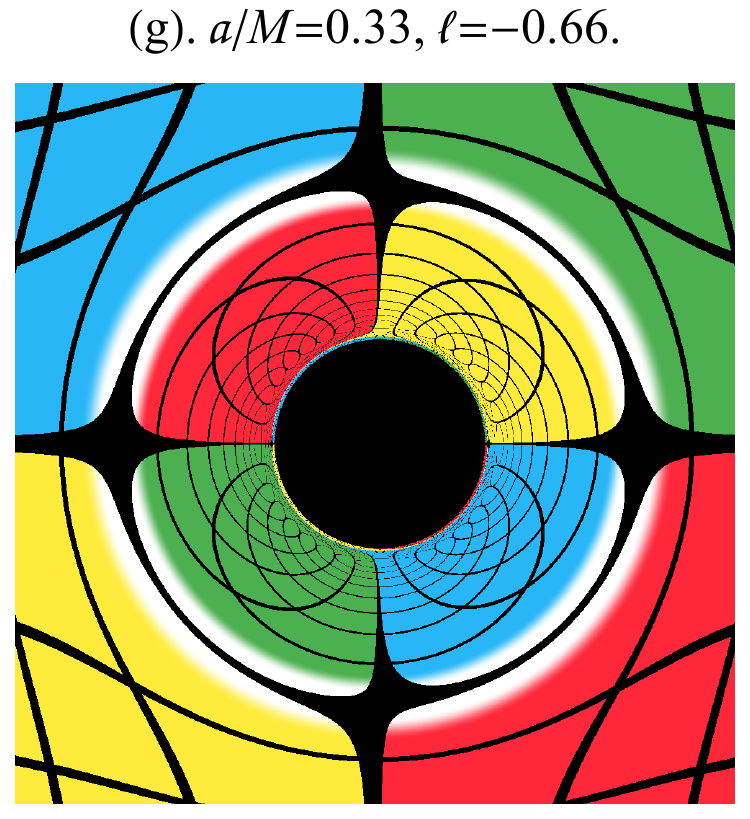}
\includegraphics[width=0.22\linewidth]{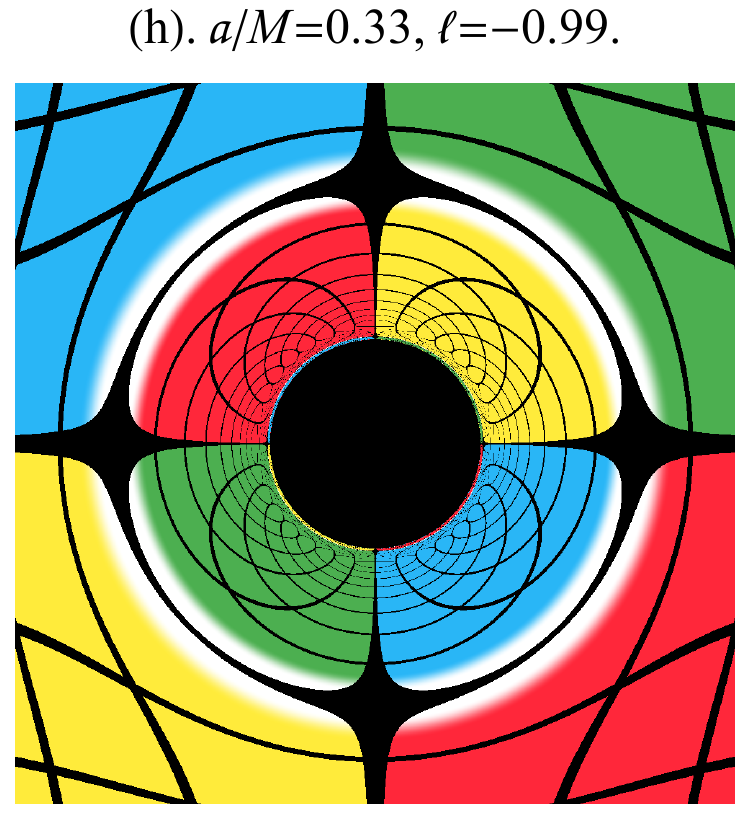}
\includegraphics[width=0.22\linewidth]{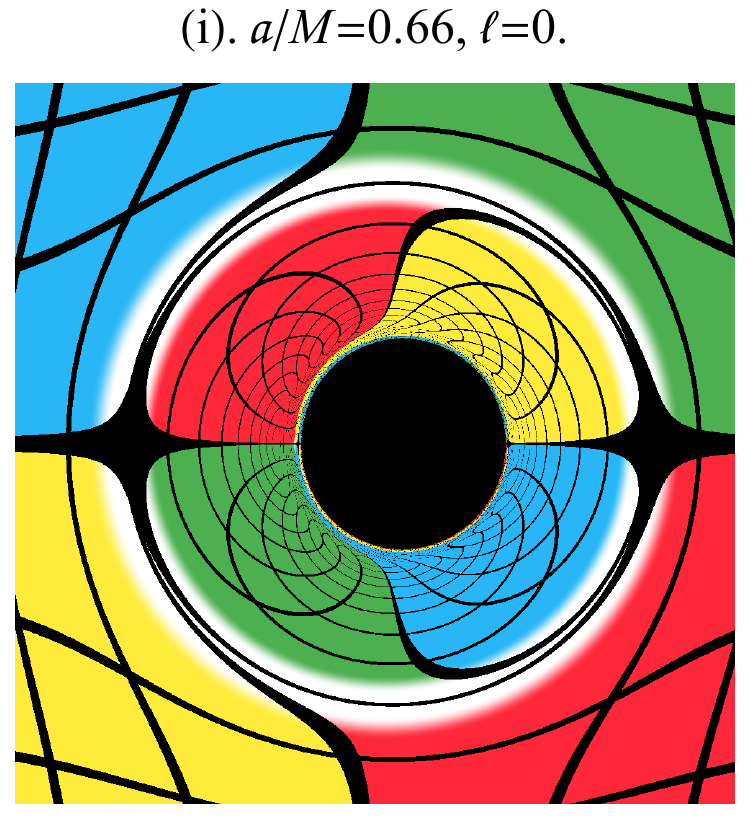}
\includegraphics[width=0.22\linewidth]{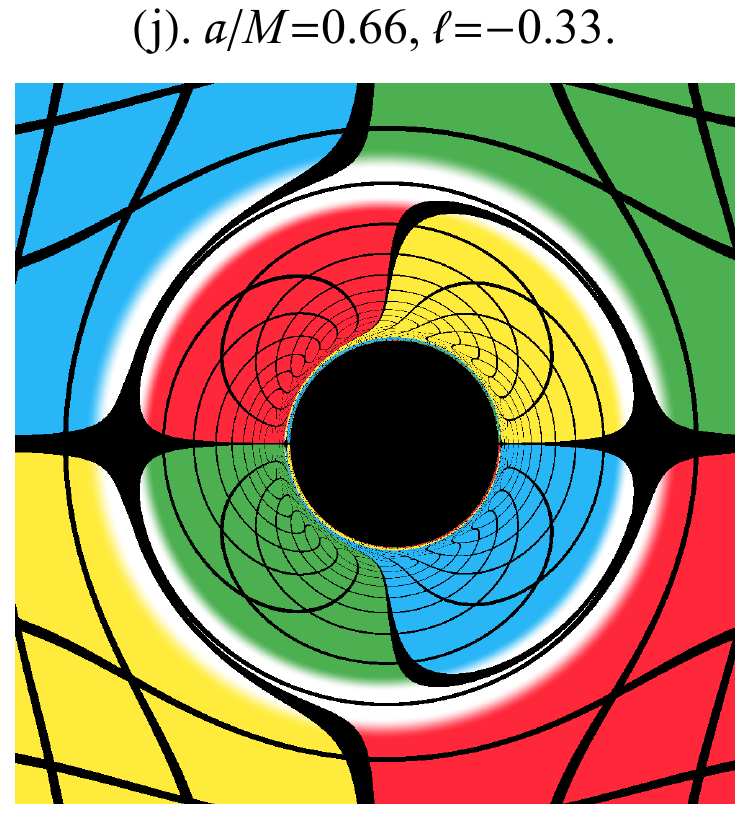}
\includegraphics[width=0.22\linewidth]{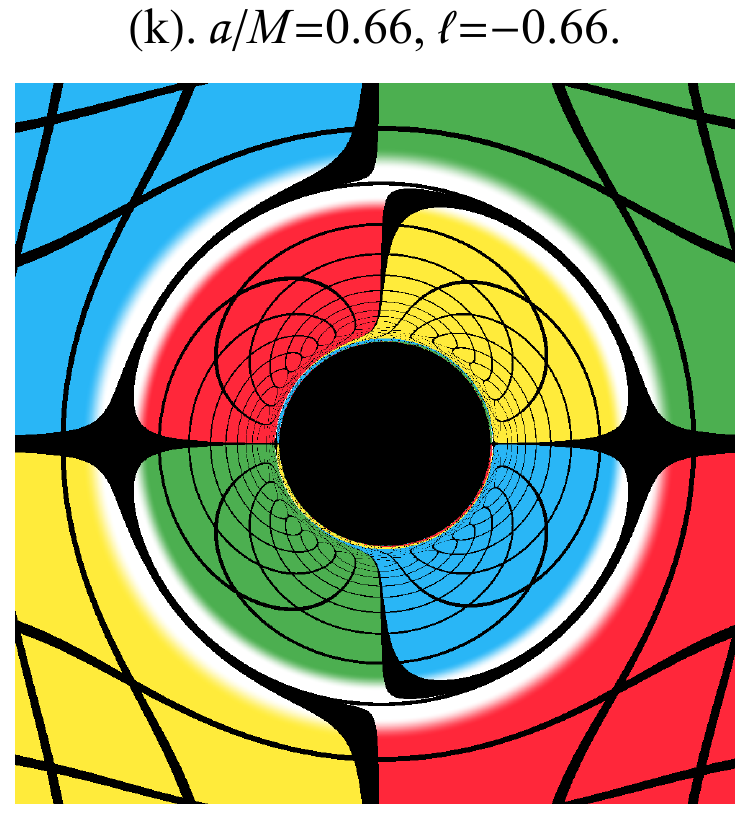}
\includegraphics[width=0.22\linewidth]{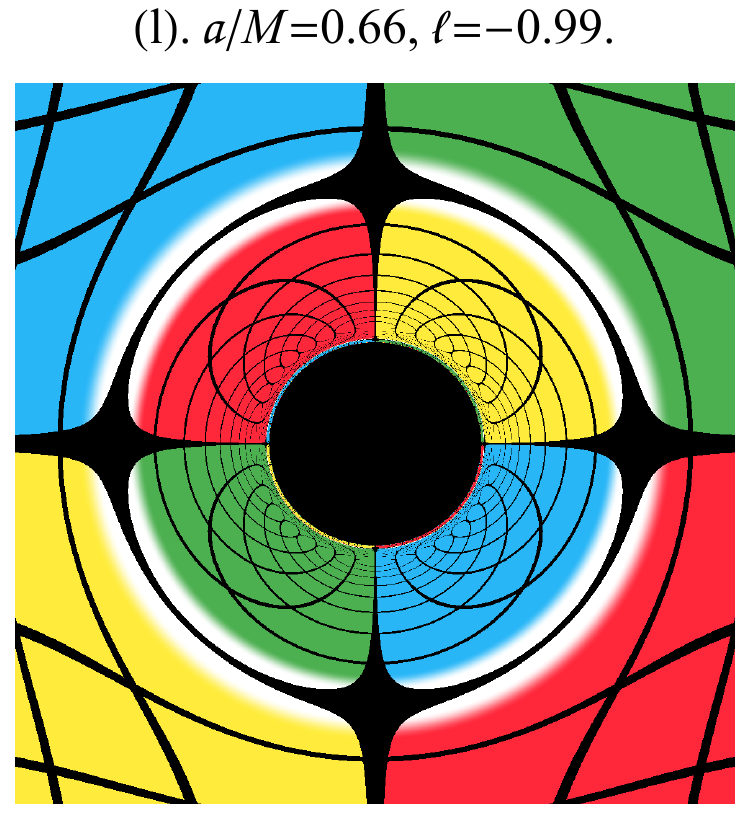}
\includegraphics[width=0.22\linewidth]{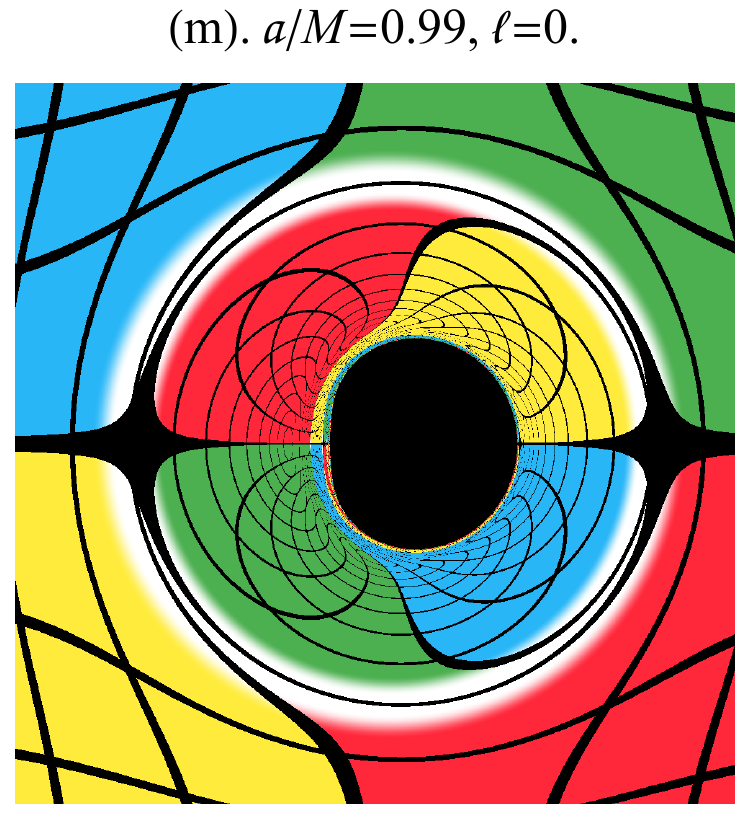}
\includegraphics[width=0.22\linewidth]{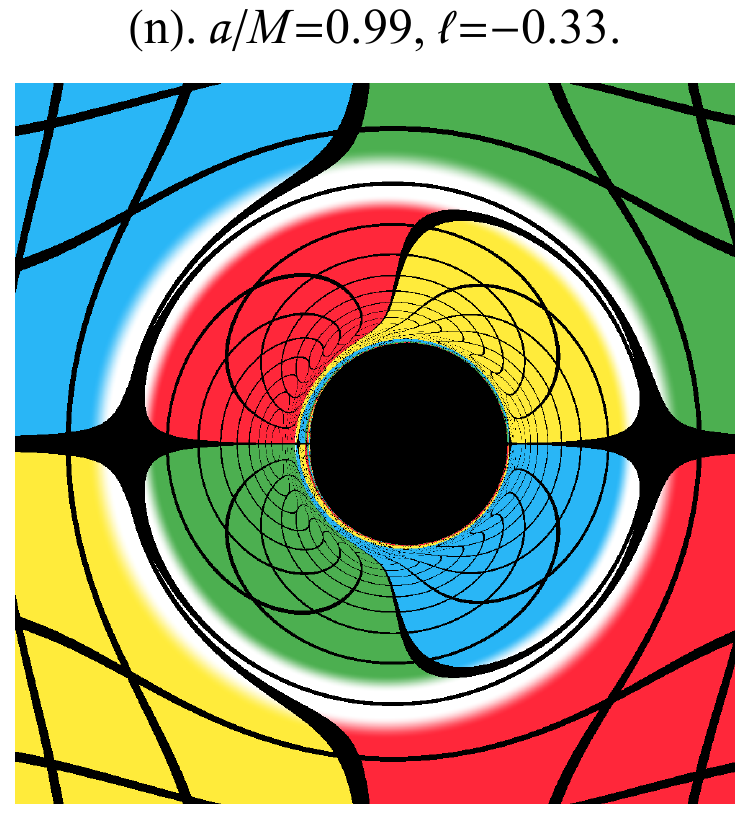}
\includegraphics[width=0.22\linewidth]{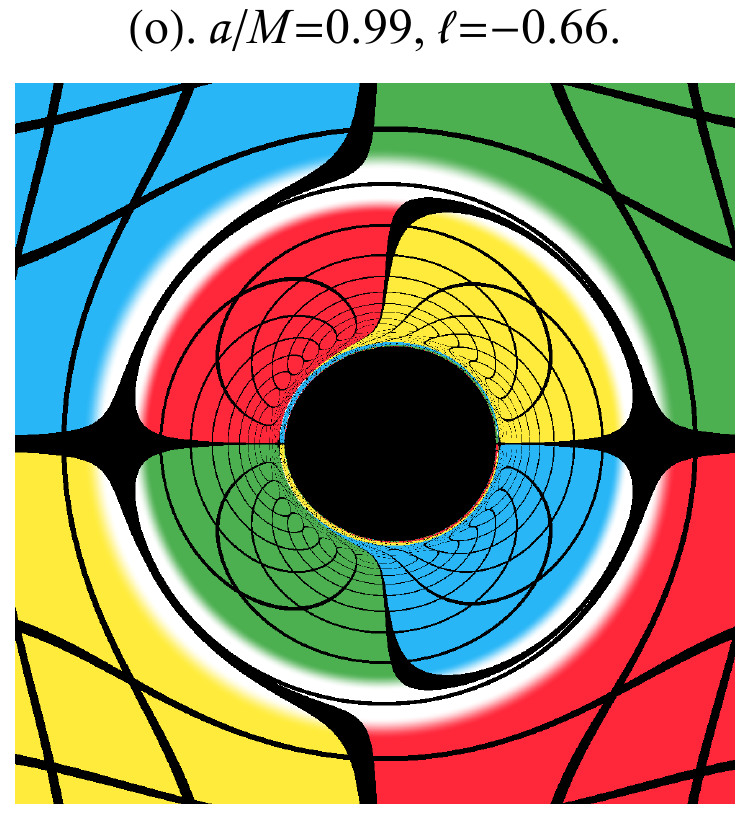}
\includegraphics[width=0.22\linewidth]{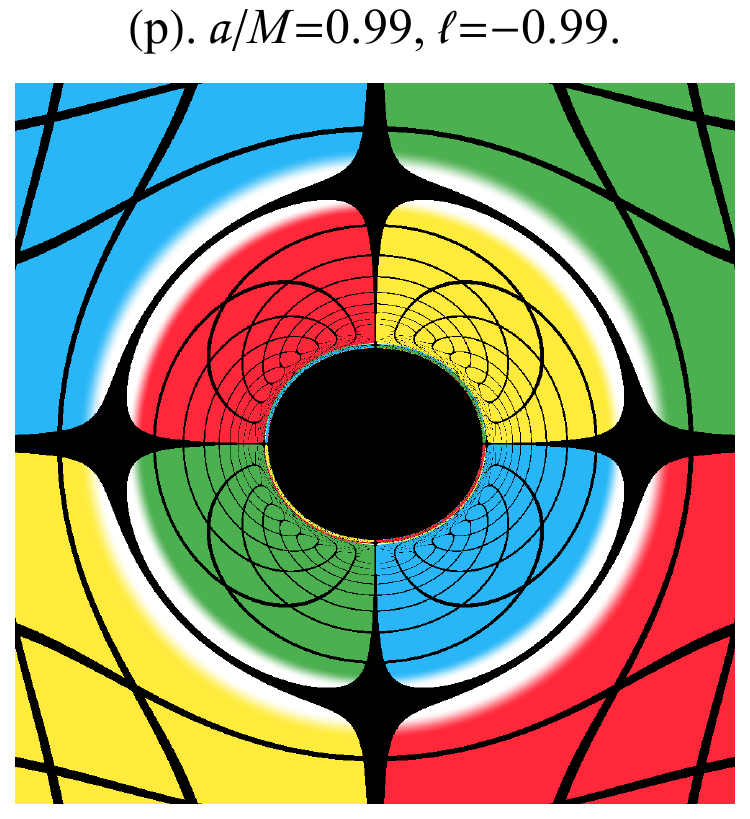}
\caption{Shadows cast by rotating BHs with a negative Lorentz violation parameter $ \ell $, as seen by an observer at $ \theta_0=\pi/2 $.}
\label{fig5}
\end{figure}

\end{widetext}

\end{document}